\documentclass[letterpaper, amsfonts, amssymb, amsmath, reprint, showkeys, nofootinbib, aps, prx, preprintnumbers, superscriptaddress,nofootinbib,longbibliography]{revtex4-2}
\usepackage[english]{babel}
\usepackage[utf8]{inputenc}
\usepackage[colorinlistoftodos, color=green!40, prependcaption]{todonotes}
\usepackage{amsthm}
\usepackage{mathtools}
\usepackage{physics}
\usepackage{xcolor}
\usepackage{graphicx}
\usepackage[left=23mm,right=13mm,top=35mm,columnsep=15pt]{geometry} 
\usepackage{adjustbox}
\usepackage{placeins}
\usepackage{caption}
\usepackage{subcaption}
\usepackage{subcaption}
\usepackage{tikz}
\usetikzlibrary{arrows.meta,positioning}
\usepackage[T1]{fontenc}
\usepackage{lipsum}
\usepackage{algorithm}
\usepackage{algpseudocode}
\usepackage{csquotes}

\usepackage[pdftex, pdftitle={Dynamical Complexity of Projected Ensembles: A Geometric Quantum State Approach}, pdfauthor={Komal Sah, Fabio Anza, Alexandra Jurgens, and James P. Crutchfield}]{hyperref} 

\begin{document}
\title{Dynamical Complexity of Projected Ensembles: A Geometric Quantum State Approach}

\author{Komal Sah}
\email{ksah@ucdavis.edu}
\affiliation{
Complexity Sciences Center and Physics and Astronomy Department,
University of California, Davis, USA
}

\author{Fabio Anza}
\email{fanza@umbc.edu}
\affiliation{
Department of Physics, University of Maryland, Baltimore County, USA
}

\author{Alexandra M. Jurgens}
\email{alexandra.jurgens@inria.fr}
\affiliation{
Inria Centre, University of Bordeaux, France
}

\author{James P. Crutchfield}
\email{chaos@ucdavis.edu}
\affiliation{
Complexity Sciences Center and Physics and Astronomy Department,
University of California, Davis, USA
}

\date{\today} 

\begin{abstract}
We study the emergence of complexity in finite-size quantum systems as their dynamics transition from closed and coherent evolution to interacting and effectively open behavior. Using a state-based geometric framework, we represent mixed quantum states as probability measures on complex projective Hilbert space. Obtained by conditioning on the environment in a fixed basis, these geometric quantum states coincide with the projected ensembles that underlie deep thermalization and emergent quantum state designs. This representation allows us to track how interactions reshape the underlying pure-state geometry. We introduce two complementary diagnostics, a distinguishability
measure based on the Wasserstein distance between nearby projected ensembles,
which quantifies sensitivity to initial states, and a state-space coverage
index, which measures long-time exploration of the subsystem state space.
Applied to the quantum kicked top, both diagnostics generally increase with
interaction strength. Their dependence on environment size is structured by
parity symmetry, with integer-spin systems often
exhibiting greater sensitivity and state-space coverage than half-integer-spin
systems. These results highlight finite-size quantum effects and provide a
geometric approach to quantifying dynamical complexity deep in the quantum
regime.
\end{abstract}

\date{\today}


\preprint{arxiv.org:2608.09705}

\maketitle

\section{Introduction}
\label{sec:introduction}

What does it mean for a dynamical system to be complex? In classical mechanics,
complexity has clear geometric interpretations: small perturbations grow,
periodic motion breaks down, competing recurrent or periodic behaviors emerge,
and phase-space structures reorganize
\cite{C1lorenz2017deterministic,C2eckmann1985ergodic,C3crutchfield1982fluctuations,C4oono1978heuristic}.
In integrable systems, motion is periodic or quasiperiodic, with trajectories
confined to invariant tori. When nonlinear interactions are introduced, these
structures deform and eventually break down, giving rise to chaotic regions
characterized by sensitive dependence on initial conditions and fractal
geometry of invariant sets
\cite{Kam1chirikov1990kam,kam2delshams1996effective}. Quantities such as the
maximal Lyapunov exponent
\cite{LLE2rosenstein1993practical,LLE1kinsner2006characterizing,LLE3Wolf1985}
and information dimension
\cite{ID1farmer1982information,ID2jurgens2021divergent} quantify this
transition by measuring trajectory instability and geometric reorganization of
invariant measures \cite{IMLasota1994}. Classical complexity can thus be
understood as the breakdown and reorganization of periodic or near-integrable phase-space
structure under interactions.

Defining complexity in quantum systems is much more subtle
\cite{QChaake1991quantum,QCzyczkowski1993generalize}. Closed quantum evolution
is linear and unitary, preserving inner products and therefore the
distinguishability between pure states. As a result, the classical notion of
state-based trajectory instability does not directly carry over. Quantum
complexity is often quantified using operator-based methods such as
out-of-time-ordered correlators (OTOCs)
\cite{OTOC1wang2025characterizing,OTOCLEchoQKT,OTOC3sieberer2019digital,OTOC4yan2020quantum},
which probe operator growth, and the Loschmidt echo
\cite{ECHO1goussev2012loschmidt,ECHO2cucchietti2003decoherence,ECHO3macri2016loschmidt},
which measures reversibility. In addition, Krylov complexity
\cite{KRYLOVrabinovici2025krylov} and entanglement entropy
\cite{EEneill2016ergodic} quantify information spreading and correlation
generation.

\begin{figure*}[htbp]
\centering
\includegraphics[width=0.99\linewidth]{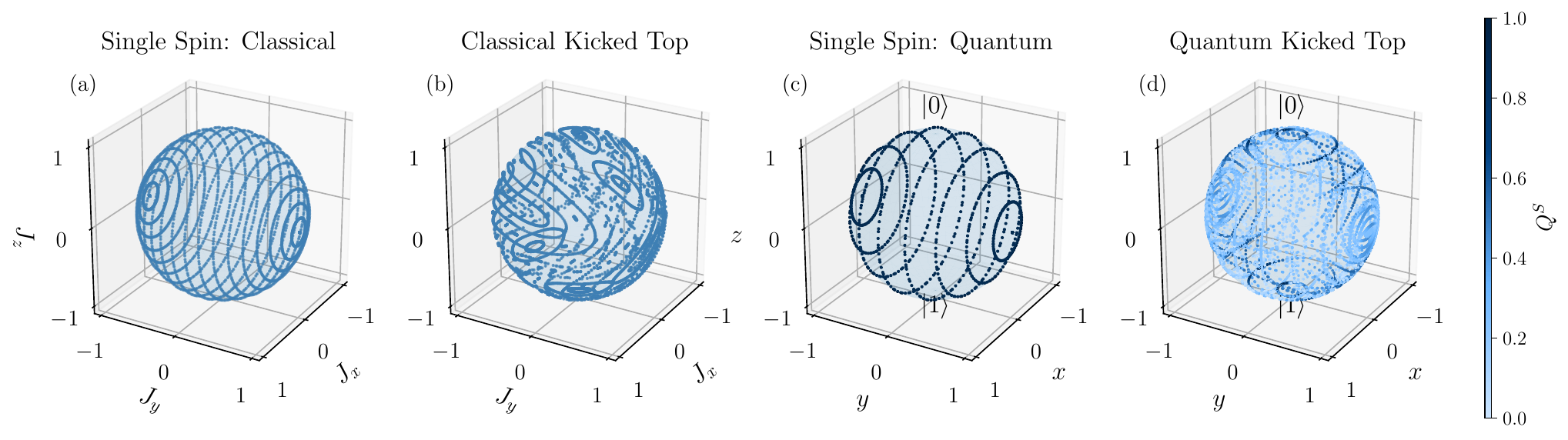}
\caption{
From periodic motion to interaction-induced complexity.
(a) Classical single-spin precession $(\kappa=0)$.
(b) Classical kicked top $(\kappa>0)$ showing trajectory deformation.
(c) Closed single-spin quantum evolution with periodic motion on $\mathbb{C}P^1$.
(d) Quantum kicked top, where interactions deform this periodic motion into an ensemble spreading over $\mathbb{C}P^1$.
}
\label{fig:CKT_QKT}
\end{figure*}

These measures capture key aspects of information scrambling and correlation
generation, but they do not describe how the geometric structure of state space
reorganizes under interactions. For a closed system, unitary motion is confined
to invariant manifolds in complex projective space $\mathbb{C}P^{d-1}$,
producing periodic or quasiperiodic dynamics analogous to integrable classical
motion (Fig.~\ref{fig:CKT_QKT}(a),(c)). Existing quantum diagnostics lack
a direct geometric analogue of classical torus-breaking or a description of
how periodic structure destabilizes once interactions are introduced.

Such a geometric counterpart naturally emerges when we consider the dynamics of an interacting quantum system that is partitioned into a system $S$ and an environment $E$. Although the joint state evolves
unitarily, interactions between $S$ and $E$ generate entanglement and render the reduced state of $S$ mixed. Decomposing the joint state in a basis of $E$ resolves this reduced state into a probability distribution over pure states on $\mathbb{C}P^{d-1}$, known as the geometric quantum state (GQS) \cite{GQS1,GQS2anza2022geometric,GQS3anza2022quantum,GQShahn2026probability}
and, in the many-body literature, as the projected ensemble
\cite{PEcotler2023emergent,PEchoi2023preparing}. The reduced dynamics of $S$ is
then the evolution of a probability measure on state space. In chaotic
many-body systems the projected ensemble can approach the uniform (Haar)
distribution over pure states, forming approximate or even exact quantum state
designs, a phenomenon termed \emph{deep thermalization}
\cite{PEcotler2023emergent,PEchoi2023preparing,PEho2022exact}. Subsequent work
has examined how this approach depends on the measurement basis, symmetries,
and conservation laws
\cite{PEippoliti2023dynamical,PEchan2024projected,PEchang2025deep,PEvarikuti2024unraveling,PEliu2026coherence},
and has realized deep thermalization experimentally
\cite{PEchoi2023preparing,PEyan2026characterizing}.

This literature has largely asked whether, and how fast, the ensemble
converges to a universal limit. We ask a complementary question about its
dynamical structure, namely how sensitively it responds to perturbations of the
initial state and how extensively it explores state space over time. To
address these questions we introduce two diagnostics whose construction is
motivated by the maximal Lyapunov exponent and the information dimension of
classical dynamics. The first is a distinguishability measure based on the
Wasserstein distance between nearby geometric quantum states. Like the maximal
Lyapunov exponent, it probes sensitivity to initial conditions, but through the
separation of probability measures on $\mathbb{C}P^{d-1}$ rather than of
individual trajectories. The second is a State-Space Coverage Index (SSCI),
defined by the Wasserstein proximity of the time-aggregated measure to the
uniform distribution. Much as the information dimension describes how an
invariant measure fills phase space, the SSCI describes how extensively the
time-aggregated measure covers state space. These diagnostics capture
finite-time instability as well as long-time, state-space-scale geometric
reorganization induced by interactions.

We illustrate the framework with the quantum kicked top, taking a single qubit
as $S$ and the remaining qubits as $E$ (Fig.~\ref{fig:CKT_QKT}(c),(d)). Without
interactions, the qubit precesses periodically on the Bloch sphere, mirroring
integrable classical motion. With interactions, entanglement deforms this
periodic evolution into a spreading ensemble on $\mathbb{C}P^1$, reflected in a
positive distinguishability measure and increased state-space coverage.
Complexity in open quantum systems thus parallels classical structural
instability at the level of evolving probability measures, and optimal
transport offers a bridge between classical dynamical intuition and open
quantum dynamics that complements operator-based diagnostics.

The following develops this framework systematically. Section~\ref{section02} contrasts global unitary invariance with reduced-state dynamics, highlighting why instability arises at the level of reduced states. It introduces the geometric quantum state representation, expressing reduced states as probability measures on complex projective space. Section~\ref{sec:section03} then develops the optimal transport geometry used to compare geometric quantum states, introducing Wasserstein distances on $\mathbb{C}P^{d-1}$. Building on this structure, Section~\ref{section04} defines two complementary diagnostics: a distinguishability measure capturing finite-time instability and a State-Space Coverage Index characterizing long-time exploration. Section~\ref{section05} applies the diagnostics to the quantum kicked top, examining their dependence on interaction strength and environment size. Finally, in Section~\ref{section06} we discuss the results and present our conclusions.

\section{Closed versus Open: Why Geometry Matters}
\label{section02}

We consider a finite-dimensional quantum system $S$ interacting with a
finite-dimensional environment $E$. Throughout, \textit{open} refers to the
reduced dynamics of a subsystem of a finite, closed system rather than to
coupling with an ideal Markovian bath. The joint many-body quantum system has Hilbert
space $\mathcal{H}_{SE}=\mathcal{H}_S \otimes \mathcal{H}_E$, where
$\dim \mathcal{H}_S = d_S$ and $\dim \mathcal{H}_E = d_E$.

The joint system evolves unitarily under the Hamiltonian
\begin{equation}
H_{SE} = H_0 + H_I,
\end{equation}
where $H_0$ generates local dynamics and $H_I$ couples $S$ and $E$. The joint
state then evolves as
\begin{equation}
|\Psi_{SE}(t)\rangle = U(t)|\Psi_{SE}(0)\rangle,
~\text{with}~ U(t)=e^{-iH_{SE}t/\hbar}.
\end{equation}

\paragraph{Global Unitary Invariance.}
The global state is represented as a vector in $\mathcal{H}_{SE}$. Consider two
nearby pure states $|\Psi_{SE}(0)\rangle$ and $|\Psi_{SE}'(0)\rangle$, whose
separation is quantified by the Fubini--Study distance
\begin{equation}
d_{FS}(|\Psi\rangle,|\Phi\rangle)
=
\arccos\!\left(|\langle\Psi|\Phi\rangle|\right).
\end{equation}
Unitary evolution preserves all inner products,
\begin{eqnarray}
\langle\Psi_{SE}(t)|\Psi_{SE}'(t)\rangle
&=&
\langle\Psi_{SE}(0)|U^\dagger U|\Psi_{SE}'(0)\rangle \nonumber\\
&=&
\langle\Psi_{SE}(0)|\Psi_{SE}'(0)\rangle,
\end{eqnarray}
so that
\begin{equation}
d_{FS}\!\left(|\Psi_{SE}(t)\rangle,|\Psi_{SE}'(t)\rangle\right)
=
d_{FS}\!\left(|\Psi_{SE}(0)\rangle,|\Psi_{SE}'(0)\rangle\right).
\end{equation}
Global distinguishability is therefore conserved for all times. Unlike
classical trajectories, which may separate exponentially under nonlinear
dynamics, the global quantum state exhibits no intrinsic trajectory
instability.

\paragraph{Loss of Invariance and the Emergence of Mixed Subsystems.}
The situation changes when we restrict attention to the subsystem $S$. Given a
product initial state $|\Psi_{SE}(0)\rangle = |\psi_S\rangle \otimes
|\psi_E\rangle$, with $|\psi_S\rangle \in \mathcal{H}_S$ and $|\psi_E\rangle
\in \mathcal{H}_E$, the interactions generated by $H_I$ produce entanglement,
so that at later times the state
\begin{equation}
|\Psi_{SE}(t)\rangle
=
\sum_{k=1}^{d_S}
\sum_{j=1}^{d_E}
\psi_{kj}(t)\,
|s_k\rangle \otimes |e_j\rangle
\label{eq:Global_state}
\end{equation}
is no longer separable. Here $\{|s_k\rangle\}$ and $\{|e_j\rangle\}$ form
orthonormal bases of $\mathcal{H}_S$ and $\mathcal{H}_E$, respectively.

The subsystem state is obtained by tracing out the environment,
\begin{equation}
\rho_S(t)
=
\mathrm{Tr}_E
\big(
|\Psi_{SE}(t)\rangle\langle\Psi_{SE}(t)|
\big).
\end{equation}
Even though the global state remains a point on $\mathbb{C}P^{d_S d_E - 1}$,
the reduced state $\rho_S(t)$ is, in general, mixed and can no longer be
described by a single pure state $|\psi_S\rangle \in \mathcal{H}_S$.

\subsection{From Pure States to Density Matrices}

The geometric object describing the dynamics changes fundamentally when we pass
from the composite system to the subsystem. Pure states of $S$ live on
$\mathbb{C}P^{d_S - 1}$, whereas mixed states are elements of the convex set
\begin{equation}
\mathcal{D}_{d_S}
=
\left\{
\rho \in \mathrm{Herm}(d_S)
\;\middle|\;
\rho \ge 0,
\ \mathrm{Tr}(\rho)=1
\right\},
\end{equation}
where $\mathrm{Herm}(d_S)$ denotes the space of $d_S\times d_S$ Hermitian
operators. Points in $\mathcal{D}_{d_S}$ represent statistical mixtures rather
than individual pure-state configurations. Thus, open-system dynamics no longer
corresponds to motion on a projective state-space manifold, but to evolution
within the density-operator space $\mathcal{D}_{d_S}$.

For initially pure subsystem states, distinguishability is measured by the
Fubini--Study distance
\begin{equation}
d_{FS}\!\left(|\psi_S\rangle,|\psi_S'\rangle\right)
=
\arccos\!\left(|\langle\psi_S|\psi_S'\rangle|\right).
\end{equation}
If the subsystem evolved unitarily, this distance would be conserved. Once
entanglement with the environment develops, however, the reduced dynamics
becomes non-unitary. Distinguishability between mixed states can then be
quantified by operator-level metrics such as the Bures distance,
\begin{equation}
d_{\mathrm B}(\rho_S,\rho_S')
=
\sqrt{\,2 - 2\sqrt{F(\rho_S,\rho_S')}\,},
\end{equation}
where $F(\rho_S,\rho_S')$ denotes the quantum fidelity,
\begin{equation}
F(\rho_S,\rho_S') =
\left[
\operatorname{Tr}
\sqrt{\sqrt{\rho_S}\,\rho_S'\sqrt{\rho_S}}
\right]^2.
\end{equation}
In general,
\begin{equation}
d_{\mathrm B}\bigl(\rho_S(t), \rho'_S(t)\bigr) \neq d_{\mathrm B}\bigl(\rho_S(0), \rho'_S(0)\bigr),
\end{equation}
so the reduced dynamics does not preserve distances.

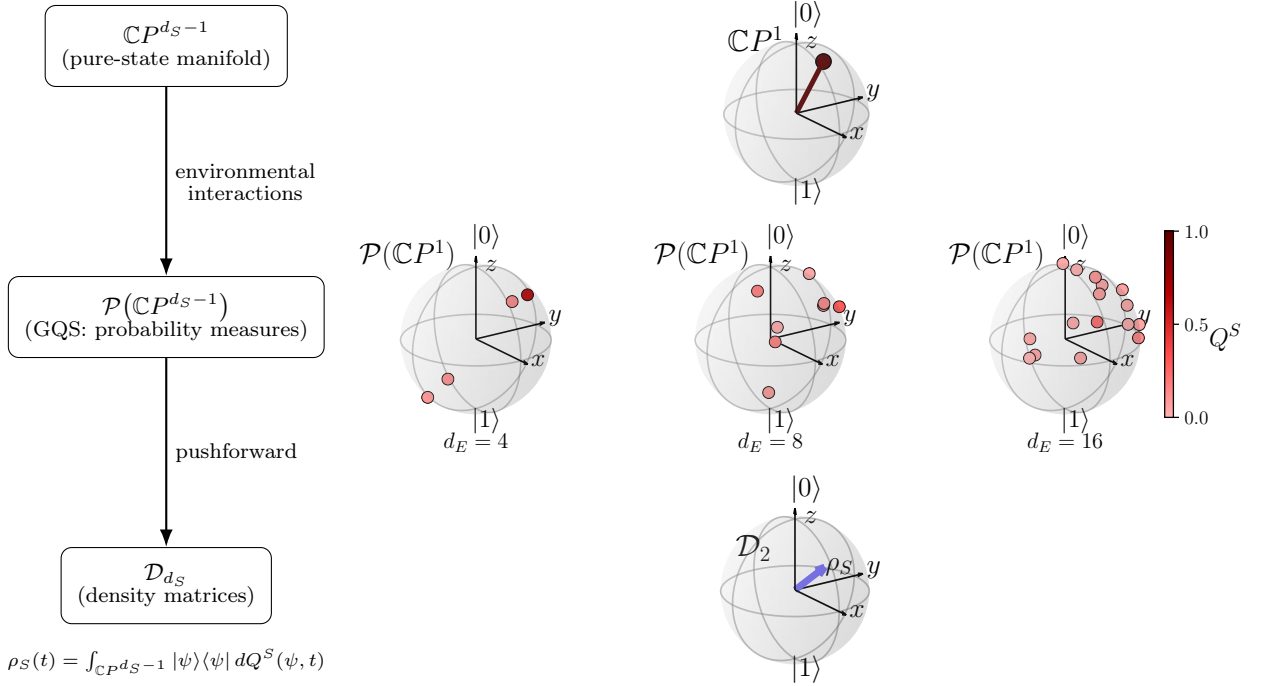
\begin{figure*}[htbp]
\centering
\begin{minipage}[c]{0.30\textwidth}
\centering
\begin{tikzpicture}[
  node distance=2.5cm,
  box/.style={draw, rounded corners, inner sep=6pt, align=center},
  arrow/.style={-Latex, thick}
]
\node[box] (cp) {$\mathbb{C}P^{d_S-1}$\\[-2pt]\footnotesize (pure-state manifold)};

\node[box, below=of cp] (P) {$\mathcal{P}\!\left(\mathbb{C}P^{d_S-1}\right)$\\[-2pt]\footnotesize (GQS: probability measures)};

\node[box, below=of P] (D) {$\mathcal{D}_{d_S}$\\[-2pt]\footnotesize (density matrices)};

\draw[arrow] (cp) -- node[right] {\footnotesize \shortstack{environmental\\interactions}} (P);
\draw[arrow] (P) -- node[right] {\footnotesize first moment} (D);

\end{tikzpicture}
\end{minipage}
\hspace{-0.02\textwidth}
\begin{minipage}[c]{0.65\textwidth}
\centering

\includegraphics[width=0.2\textwidth]{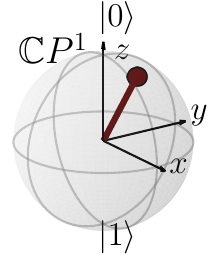}

\vspace{0.5em}

\includegraphics[width=1.0\textwidth]{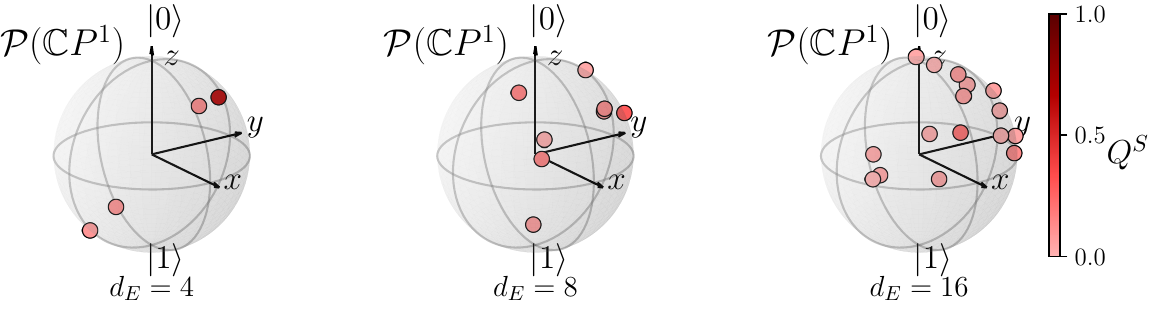}

\vspace{0.5em}

\includegraphics[width=0.2\textwidth]{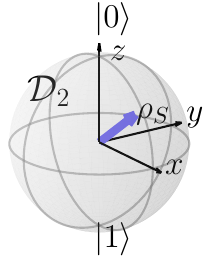}

\end{minipage}

\caption{
Left: Relation between pure states, GQS, and density matrices. Right: For a single-qubit subsystem, interactions with different environment sizes ($L_E=2,3,4$, $d_E=2^{L_E}$) produce distinct ensembles on $\mathbb{C}P^1$ with the same reduced density matrix $\rho_S$, showing that $\rho_S$ does not capture the geometric structure induced by environment conditioning.}
\label{fig:gqs_pushforward}
\end{figure*}

\subsection{Limitations of the Density Matrix Description}
The passage from the smooth projective Hilbert space of pure states to the
operator space of density matrices makes it difficult to translate classical,
state-based complexity measures directly.

Every point on $\mathbb{C}P^{d_S-1}$ corresponds to a unique pure quantum
state. In contrast, points in $\mathcal{D}_{d_S}$ do not admit such a
one-to-one interpretation. A mixed density matrix admits infinitely many convex
decompositions,
\begin{equation}
\rho_S
=
\sum_j \lambda_j
|\psi_j\rangle\langle\psi_j|,
\qquad
\lambda_j \ge 0,
\quad
\sum_j \lambda_j = 1,
\end{equation}
and these distinct ensembles generally correspond to different probability
measures on $\mathbb{C}P^{d_S-1}$. All of them, however, map to the same
operator $\rho_S$. The space $\mathcal{D}_{d_S}$ thus identifies many
geometrically distinct probability distributions on $\mathbb{C}P^{d_S-1}$ as a
single point.

Consequently, operator-level distances such as the Bures distance compare
density matrices only through their averaged statistical effect and are
insensitive to how probability mass is arranged across $\mathbb{C}P^{d_S-1}$.
Two density matrices may be arbitrarily close in Bures distance, or even
identical, while their underlying pure-state components occupy widely separated
regions of projective space. This degeneracy is significant when formulating a
quantum analogue of classical trajectory instability. To track how quantum
ensembles spread, reorganize, or develop geometric structure, working solely
within $\mathcal{D}_{d_S}$ is insufficient.

\subsection{From Density Matrices to Geometric Quantum States}

A natural ensemble structure emerges directly from the global system-environment state.
Regrouping the expansion in Eq.~(\ref{eq:Global_state}) by environment basis state yields the environment-conditioned decomposition
\begin{equation}
|\Psi_{SE}(t)\rangle = \sum_{j=1}^{d_E} \sqrt{\lambda_j^{E}(t)} \,|\chi_j^{S}(t)\rangle |e_j\rangle .
\end{equation}
Throughout this work, $\{|e_j\rangle\}$ is taken to be the computational basis.
For an environment of $L_E$ qubits, with $d_E = 2^{L_E}$, this is the product basis
\[
\{|e_j\rangle\}_{j=1}^{2^{L_E}} = \{|b_1 b_2 \cdots b_{L_E}\rangle \mid b_\ell \in \{0,1\}\},
\]
where $j$ simply enumerates the $2^{L_E}$ binary strings.
For example, when $d_E=4$,
\[
\{|e_j\rangle\}_{j=1}^{4} = \{|00\rangle, |01\rangle, |10\rangle, |11\rangle\}.
\]
Although the resulting geometric quantum states depend on this choice of conditioning, the computational basis provides a natural reference for qubit systems. Such basis dependence is intrinsic to projected ensembles and has been studied in detail in Refs.~\cite{PEippoliti2023dynamical,PEvarikuti2024unraveling,PEliu2026coherence}.

The conditional subsystem states and their probabilities are
\begin{equation}
|\chi_j^{S}(t)\rangle = \frac{1}{\sqrt{\lambda_j^{E}(t)}} \sum_{k=1}^{d_S} \psi_{kj}(t)|s_k\rangle,
\label{eq:chi}
\end{equation}
\begin{equation}
\lambda_j^{E}(t) = \sum_{k=1}^{d_S} |\psi_{kj}(t)|^2 .
\end{equation}
Terms with $\lambda_j^E(t)=0$ are omitted, since the corresponding conditional states in Eq.~\eqref{eq:chi} are undefined.
The ensemble $\{\lambda_j^{E}(t),|\chi_j^{S}(t)\rangle\}_j$ therefore defines a probability distribution over $\mathbb{C}P^{d_S-1}$.
Interactions with the environment redistribute probability weight among its components, reshaping the geometric organization of the subsystem ensemble on projective space (Fig.~\ref{fig:gqs_pushforward}).
Tracing over the environment yields the reduced density matrix
\begin{equation}
\rho_S(t) = \sum_{j=1}^{d_E} \lambda_j^{E}(t) |\chi_j^{S}(t)\rangle\langle\chi_j^{S}(t)|.
\end{equation}

While $\rho_S(t)$ reproduces all observable statistics of the subsystem, it encodes this ensemble structure only implicitly.
As discussed above, many distinct ensembles share the same density matrix, so the operator description does not retain how environment conditioning arranges probability mass on $\mathbb{C}P^{d_S-1}$. The projected ensemble can undergo qualitative changes while the reduced density matrix remains maximally mixed \cite{PEliu2026coherence}.

For this reason we lift the reduced dynamics to the space $\mathcal{P}\!\left(\mathbb{C}P^{d_S-1}\right)$ of probability measures on projective Hilbert space.
Each conditional state $|\chi_j^{S}(t)\rangle$ defines a point
\begin{equation}
\mathbf{Z}_j^{S}(t) = \big[\psi_{1j}(t):\cdots:\psi_{d_S j}(t)\big] \in \mathbb{C}P^{d_S-1},
\end{equation}
since the normalization in Eq.~\eqref{eq:chi} drops out of the homogeneous coordinates.
The subsystem state is then represented as the geometric quantum state (GQS)
\begin{equation}
Q^S(Z,t) = \sum_{j=1}^{d_E} \lambda_j^{E}(t) \, \delta^Z_{\mathbf{Z}_j^{S}(t)} \in \mathcal{P}\!\left(\mathbb{C}P^{d_S-1}\right).
\end{equation}
For a fixed environment basis, $Q^S$ is precisely the projected ensemble of Refs.~\cite{PEcotler2023emergent,PEho2022exact,PEippoliti2023dynamical,PEchan2024projected,PEanza2026nonequilibrium}. We retain the GQS formulation because it treats the ensemble explicitly as a probability measure on $\mathbb{C}P^{d_S-1}$, which helps with the optimal-transport comparisons developed below. If the subsystem and environment remain unentangled, all conditional states coincide and the GQS reduces to a single point mass
\begin{equation}
Q^S(Z,t) = \delta^Z_{\mathbf{Z}^{S}(t)},
\label{eq:Pure_dist}
\end{equation}
corresponding to a pure state in $\mathbb{C}P^{d_S-1}$.

Fig.~\ref{fig:gqs_pushforward} illustrates this distinction for a single qubit.
In the density-matrix picture, a mixed state corresponds to a single point inside the Bloch ball.
In the GQS representation, the same state is a probability distribution over pure states on the Bloch sphere ($\mathbb{C}P^1$).
The density matrix is recovered as the first moment of the GQS,
\begin{equation}
\rho_S(t)= \int_{\mathbb{C}P^{d_S-1}} |\psi\rangle\langle\psi| \, dQ^S(\psi,t).
\end{equation}

In this framework, instability is formulated not through operator differences in $\mathcal{D}_{d_S}$ but through the evolution of probability mass on $\mathbb{C}P^{d_S-1}$.
Because $\mathbb{C}P^{d_S-1}$ is a smooth Riemannian manifold equipped with the Fubini--Study metric, transport-based distances can be defined directly between subsystem ensembles, restoring a geometric description of subsystem dynamics analogous to that of classical dynamical systems.

\section{Optimal Transport Between Projected Ensembles}
\label{sec:section03}
Subsystem states are represented as probability measures on the projective Hilbert space $\mathbb{C}P^{d_S-1}$ within the geometric quantum state framework (Sec.~\ref{section02}).
To quantify distances between such measures, we employ the Wasserstein distance from optimal transport theory \cite{villani2009optimal}.

Quantum extensions of optimal transport and related distance-based approaches have been developed in operator settings, including formulations on density matrices and quantum channels \cite{W1friedland2022quantum,W2camacho2025critical,W3de2021quantum,PDwang2021quantum}.
The GQS formulation adopted here instead works at the level of probability measures over pure states.
This allows us to use classical optimal transport definitions and solvers, with the Fubini--Study distance providing the natural cost on projective Hilbert space. Wasserstein distances have also recently been advocated as a measure of convergence in deep thermalization, applied to the distributions of observables over the projected ensemble \cite{PEbejan2025matchgate}.
Here, $W_1$ is instead defined directly between measures on $\mathbb{C}P^{d_S-1}$ and used dynamically, to track how nearby ensembles separate in time.

Consider two geometric quantum states
\begin{eqnarray}
Q^{S}(Z)&=& \sum_{i} \lambda^{S}_i  \delta^Z_{\mathbf{Z}^{S}_i},\\
Q^{S'}(Z)&=& \sum_{j} \lambda^{S'}_j\delta^Z_{\mathbf{Z}^{S'}_j}.
\end{eqnarray}
The Wasserstein distance of order $p$ between these measures is
\begin{equation}
W_p(Q^{S},Q^{S'}) = \left( \inf_{\pi\in\Pi(Q^{S},Q^{S'})} \sum_{i,j}\pi_{ij}\, d_{FS}\!\left(\mathbf{Z}_i^{S},\mathbf{Z}_j^{S'}\right)^p \right)^{1/p},
\end{equation}
where $d_{FS}$ is the Fubini--Study distance on $\mathbb{C}P^{d_S-1}$ and $\Pi(Q^{S},Q^{S'})$ is the set of transport plans, or couplings, $\pi_{ij}$ satisfying
\begin{eqnarray}
\pi_{ij} &\geq& 0,\\
\sum_{j} \pi_{i j} &=& \lambda^{S}_i ,\\
\sum_{i} \pi_{ij} &=& \lambda^{S'}_j .
\end{eqnarray}
The Wasserstein distance is thus the minimal cost of transporting probability mass from one geometric quantum state to the other, with the cost set by the Fubini--Study distance between pure states.
The Bures distance compares only the first moments of two projected ensembles, namely their reduced density matrices, and is therefore insensitive to how probability mass is arranged within each ensemble. The Wasserstein distance compares the projected ensembles themselves and thus captures rearrangements of subsystem ensembles on $\mathbb{C}P^{d_S-1}$ induced by interactions with the environment, including those that leave $\rho_S$ unchanged.

In the following we focus on $p=1$ and refer to $W_1(Q^{S},Q^{S'})$ simply as the Wasserstein distance.
For pure states it reduces to the Fubini--Study distance,
\begin{equation}
W_1\!\left(\delta_{\psi},\delta_{\phi}\right) = d_{FS}(|\psi\rangle,|\phi\rangle),
\label{eq:W1_dFS}
\end{equation}
so a single metric describes subsystem states throughout their transition from pure to mixed under interactions with the environment.
Details of the discrete formulation and numerical implementation are provided in Appendix~\ref{app:wasserstein}.
The following section uses this distance to define a dynamical diagnostic of complexity based on how nearby geometric quantum states evolve under interactions.
\section{Ensemble-Based Diagnostics of Dynamical Complexity}
\label{section04}
\subsection{Distinguishability Measure}

Sensitivity to initial conditions is a defining feature of complexity in classical dynamical systems, typically quantified by the maximal Lyapunov exponent, which measures the exponential divergence of nearby trajectories in phase space.

In open quantum systems, subsystem states do not follow single trajectories.
Interactions instead spread an initially localized state into a GQS, a probability measure on $\mathbb{C}P^{d_S-1}$ (Sec.~\ref{section02}).
This motivates replacing trajectory-based diagnostics with an ensemble-based notion of sensitivity, in which the Wasserstein distance (Sec.~\ref{sec:section03}) quantifies how nearby subsystem ensembles separate under identical dynamics.

Let $Q^S$ denote a reference GQS, and let $\{Q_m^{S'}\}_{m=1}^{M}$ be a collection of nearby perturbed states characterized by the perturbation scale $\epsilon$, such that $Q_m^{S'}\to Q^S$ as $\epsilon\to0$.
For each perturbation, the separation between the evolved perturbed and reference states is
\begin{equation}
W_{1,m}(t) = W_1\!\left( Q_m^{S'}(Z,t), Q^S(Z,t) \right),
\end{equation}
and the distinguishability measure is
\begin{equation}
\Gamma = \frac{1}{T} \sum_{t=0}^{T-1} \left\langle \ln\!\left[ \frac{W_{1,m}(t)}{W_{1,m}(0)} \right] \right\rangle_m ,
\label{LCE4}
\end{equation}
where $\langle \cdot \rangle_m$ denotes an average over the $M$ perturbations.
This average reduces dependence on a particular perturbation and captures the typical distinguishability growth near the reference GQS.
The averaging time $T$ is chosen according to the numerically estimated recurrence timescale when recurrences are present, and is otherwise taken sufficiently large to capture the long-time behavior of $\Gamma$.

The quantity $\Gamma$ measures the average logarithmic growth of the Wasserstein distance between initially nearby subsystem ensembles, and in this sense acts as a Lyapunov-type diagnostic on $\mathcal{P}(\mathbb{C}P^{d_S-1})$.
Without interactions, the subsystem evolves unitarily, its states remain pure, and the Wasserstein distance reduces to the Fubini--Study distance, Eq.~\eqref{eq:W1_dFS}, so separations are preserved and $\Gamma=0$.
With interactions, the reduced dynamics may either amplify or contract ensemble separations. This is consistent with the exact transport description of projected-ensemble dynamics \cite{PEanza2026nonequilibrium}. When $S$ evolves independently of $E$, the ensemble is carried rigidly by a volume-preserving (Liouvillian) flow on $\mathbb{C}P^{d_S-1}$, so a nonzero $\Gamma$ requires the environment-induced redistribution of conditional weights and the compressible flow identified in that description.

A positive $\Gamma$ signals sensitivity and complex dynamics, $\Gamma=0$ corresponds to preserved distinguishability, as in the noninteracting periodic case, and $\Gamma<0$ indicates average contraction over the observation window.
Ensemble distinguishability thus provides a notion of sensitivity for open quantum systems in the spirit of classical Lyapunov analysis.

\subsection{State-Space Coverage Index}

While the distinguishability measure $\Gamma$ captures finite-time sensitivity, it does not quantify how extensively state space is explored.
In classical systems, sensitivity and global organization are distinct, because trajectories may remain confined to invariant structures such as tori or spread across larger regions of phase space as these structures break down.
This long-time spread is often quantified by geometric measures such as the information dimension \cite{ID1farmer1982information,ID2jurgens2021divergent} (see Appendix~\ref{app:information_dimension}).
For subsystem dynamics described by probability measures on $\mathbb{C}P^{d_S-1}$, the analogous question is how far interactions redistribute probability mass across state space, and we introduce a geometric diagnostic to address it.

For each time $t$, the GQS $Q^{S}(\cdot,t)$ is a probability measure in $\mathcal{P}(\mathbb{C}P^{d_S-1})$.
We define the time-aggregated GQS as
\begin{equation}
\nu^{S}(A) = \frac{1}{T_{\mathcal{S}}} \sum_{t=0}^{T_{\mathcal{S}}-1} Q^{S}(A,t), \qquad A \subset \mathbb{C}P^{d_S-1},
\end{equation}
which represents the aggregate distribution of subsystem states over $T_{\mathcal{S}}$ time steps.
Numerically, $T_{\mathcal{S}}$ is taken sufficiently large that $\nu^{S}$ approximates the long-time aggregate distribution, formally corresponding to the limit $T_{\mathcal{S}}\to\infty$ when it exists.

The State-Space Coverage Index (SSCI) is then
\begin{equation}
\mathcal{S}_p = 1 - \frac{ W_p(\nu^{S},\sigma) }{ W_p(\delta^Z_{Z_0},\sigma) },
\end{equation}
where $W_p$ is the $p$-Wasserstein distance and $\sigma$ is the uniform measure on $\mathbb{C}P^{d_S-1}$ induced by the Fubini--Study geometry, which coincides with the Haar measure on pure states.
The measure $\sigma$ represents uniform coverage, in which no pure state is preferred, while the Dirac measure $\delta^Z_{Z_0}$ represents maximal localization on a single pure state $Z_0$.
The denominator is therefore the distance between maximally localized and uniformly spread distributions.
Because $\mathbb{C}P^{d_S-1}$ is homogeneous under unitary transformations, this distance does not depend on the choice of $Z_0$ and provides a fixed normalization scale.
Throughout the numerical analysis we take $p=1$ and report $\mathcal{S}_1$, which measures coverage with the same Wasserstein geometry used for the distinguishability measure.

\begin{figure}[htbp]
\centering
\includegraphics[width=0.75 \linewidth]{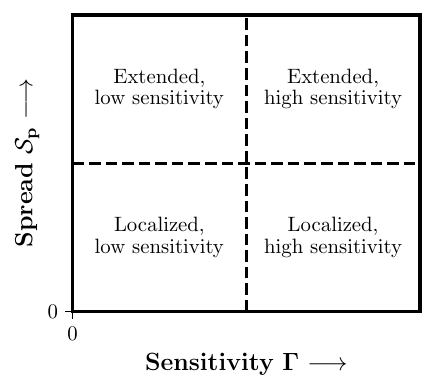}
\caption{
Schematic classification of subsystem dynamics in the nonnegative-$\Gamma$ regime using the pair $(\Gamma,\mathcal{S}_p)$. Increasing $\Gamma$ indicates greater sensitivity to initial states, while increasing $\mathcal{S}_p$ indicates broader long-time exploration of state space. The four quadrants distinguish localized low-sensitivity, extended low-sensitivity, localized high-sensitivity, and extended high-sensitivity dynamics.
}
\label{fig:gamma_sp_regime_diagram}
\end{figure}

By construction, $\mathcal{S}_p=1$ if and only if $\nu^{S}=\sigma$, corresponding to uniform exploration of state space.
In contrast, $\mathcal{S}_p=0$ corresponds to maximal localization, where the distribution remains concentrated on a single pure state throughout the evolution.
Intermediate values indicate progressively broader long-time spread across $\mathbb{C}P^{d_S-1}$.
Thus, $\mathcal{S}_p$ captures the global geometric footprint of subsystem dynamics by quantifying how broadly probability mass is distributed over projective Hilbert space, independently of local separation rates.

$\mathcal{S}_p$ differs from standard diagnostics of deep thermalization, which compare the \emph{instantaneous} ensemble $Q^S(\cdot,t)$ with the Haar measure through its statistical moments \cite{PEcotler2023emergent,PEippoliti2023dynamical,PEchan2024projected}.
Because $\nu^S$ aggregates over time, $\mathcal{S}_p$ is more closely related
to Hilbert-space ergodicity of temporal ensembles
\cite{PEmark2024maximum,PEpilatowsky2024hilbert}, which periodic driving
obstructs \cite{PEpilatowsky2024hilbert} but environment conditioning can
circumvent at the level of the subsystem ensemble.

The pair $(\Gamma,\mathcal{S}_p)$ provides a two-dimensional characterization of subsystem dynamics by separating sensitivity to initial states from long-time state-space exploration, as summarized schematically in Fig.~\ref{fig:gamma_sp_regime_diagram} for the nonnegative-$\Gamma$ regime.
Small $\Gamma$ and small $\mathcal{S}_p$ indicate localized dynamics with weak sensitivity, as in periodic or quasiperiodic motion confined to a small region of $\mathbb{C}P^{d_S-1}$.
Small $\Gamma$ but large $\mathcal{S}_p$ indicates weakly sensitive dynamics that explore an extended region of state space.
Larger $\Gamma$ corresponds to greater sensitivity, which remains confined to a small region of state space when $\mathcal{S}_p$ is small and is accompanied by broad exploration when $\mathcal{S}_p$ is large.

The State-Space Coverage Index thus complements the distinguishability measure by quantifying the extent of exploration, enabling a geometric interpretation of subsystem dynamics in terms of both sensitivity and spread. All values reported below are computed by conditioning in the computational environment basis and are therefore basis dependent.

\section{Sensitivity and State-Space Exploration in the Quantum Kicked Top}
\label{section05}

We apply the diagnostics of Sec.~\ref{section04} to the quantum kicked top. The kicked top is a foundational model for studying the transition from periodic to complex behavior \cite{QKTTlombardi2011entanglement}. It has a classically chaotic counterpart and has been realized in several experimental platforms \cite{EXPTchaudhury2009quantum,EEneill2016ergodic,EXPTkrithika2019nmr,EXPTanand2024simulating}, which makes it a natural test case.

For a system of $L=2j$ qubits with total spin $j$, the kicked top is described by the time-dependent Hamiltonian
\begin{equation}
H(t) = \frac{\pi}{2\tau} J_y
+ \frac{\kappa}{2j} J_z^2 \sum_{n\in \mathbb{Z}}\delta(t-n\tau),
\end{equation}
where the first term generates a rotation about the $y$-axis and the second term introduces nonlinear interactions between the qubits. Throughout this work we set $\hbar=1$ and $\tau=1$, so time is measured in units of the kick period, with $t_n=n$ denoting the number of Floquet kicks. With $J_y = \frac{1}{2}\sum_{i=1}^{L} \sigma_y^{(i)}$ and $J_z = \frac{1}{2}\sum_{i=1}^{L} \sigma_z^{(i)}$, the nonlinear term becomes
\begin{equation}
    J_z^2 = \frac{1}{4}\left( L\, I + 2\sum_{i<k}\sigma_z^{(i)} \sigma_z^{(k)} \right).
\end{equation}
The identity term contributes only an overall phase, while the remaining terms describe homogeneous pairwise interactions between all qubits, with collective coupling strength set by $\kappa$. For $L=3$, this all-to-all interaction graph is equivalent to a three-site ring with periodic nearest-neighbor couplings. For $L>3$, the collective $J_z^2$ term also couples non-nearest-neighbor qubits, introducing additional interaction pathways beyond those of the three-qubit system.

The subsystem and environment dimensions introduced earlier are $d_S = 2^{L_S}$ and $d_E = 2^{L_E}$, with $L = L_S + L_E$. The Floquet operator
\begin{equation}
    U = e^{-i\kappa J_z^2/2j}\,e^{-i\pi J_y/2}
\end{equation}
governs the discrete-time evolution and drives the system away from product structure by generating multipartite entanglement.

We initialize the system in a spin-coherent product state,
\begin{align}
|\psi(\theta,\phi)\rangle
& = \cos\!\left(\tfrac{\theta}{2}\right)|0\rangle
+ e^{i\phi}\sin\!\left(\tfrac{\theta}{2}\right)|1\rangle, \\
|\Psi_{SE}(0)\rangle
& = \bigotimes_{\ell=1}^{L}|\psi(\theta,\phi)\rangle,
\end{align}
and focus on the reduced dynamics of a single-qubit subsystem ($L_S=1$, $d_S=2$), represented by its GQS in the computational environment basis (Sec.~\ref{section02}). Initially, this GQS is localized,
\begin{equation}
Q^S(Z,0)= \delta^Z_{\mathbf{Z}^{S}},
\end{equation}
where $\mathbf{Z}^{S}\in\mathbb{C}P^{1}$ is the point on the Bloch sphere corresponding to $|\psi(\theta,\phi)\rangle$.

\begin{figure*}
    \centering
    \includegraphics[width=1.\linewidth]{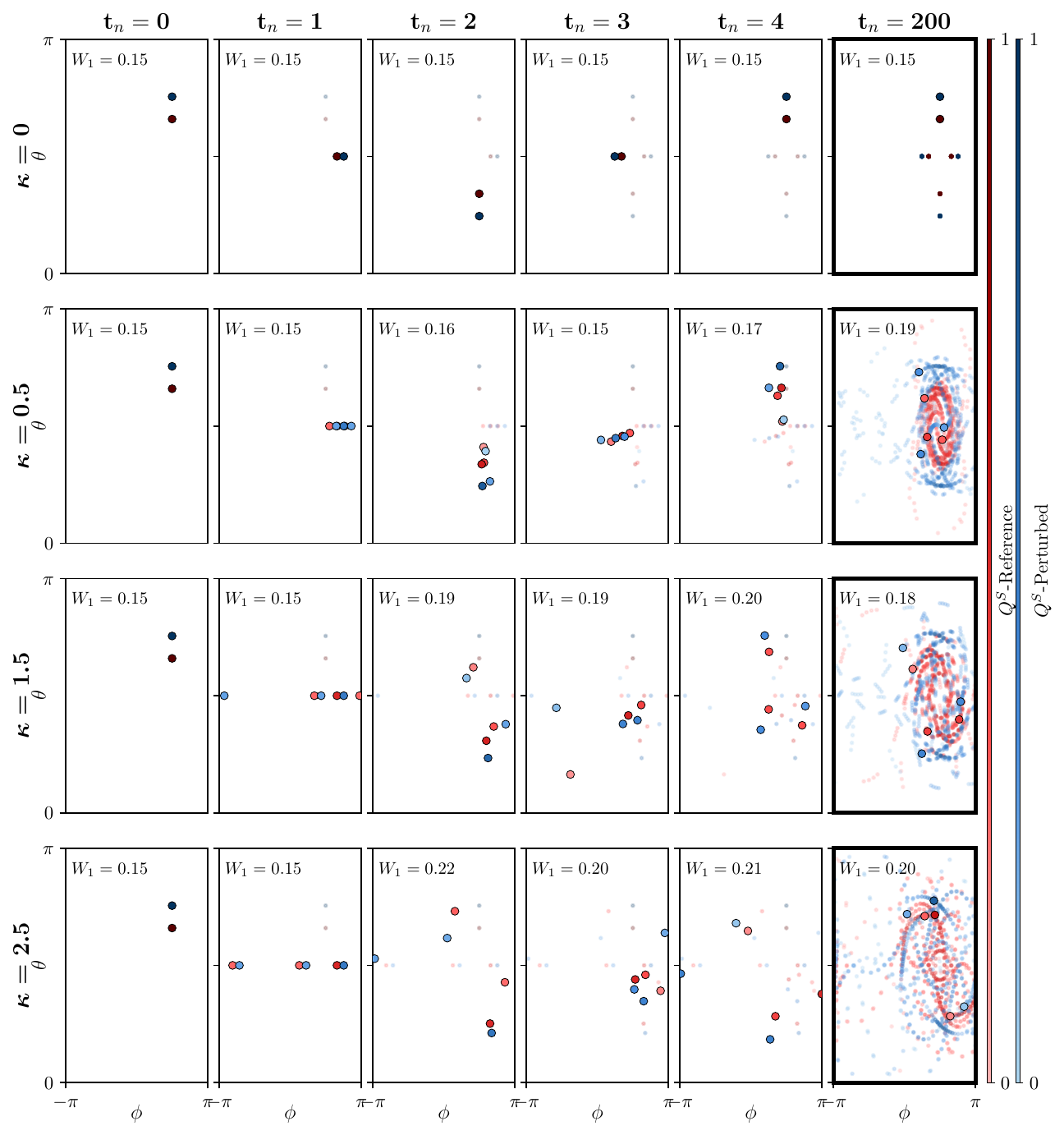}
    \caption{Evolution of two neighboring geometric quantum states in the three-qubit quantum kicked top. Columns show the projected ensembles (GQSs) on $\mathbb{C}P^{1}$ after $t_n=0,1,2,3,4,$ and $200$ Floquet kicks, while rows correspond to interaction strengths $\kappa$. The separation $W_1(t_n)=d_{FS}$ is preserved for noninteracting dynamics ($\kappa=0$). For $\kappa>0$, $W_1(t_n)$ varies in time, indicating loss of invariance as the ensembles deform and spread across state space, with stronger effects at larger $\kappa$.}
    \label{fig:Int_Deform}
\end{figure*}

To probe sensitivity to initial conditions, we perturb the initial spin-coherent state using small isotropic random rotations with Gaussian-distributed rotation angles of standard deviation $\epsilon$. This generates an ensemble of nearby product states,
\begin{equation}
|\Psi_{SE,m}'(0)\rangle
=
\bigotimes_{\ell=1}^{L}
|\psi(\theta_m',\phi_m')\rangle,
\end{equation}
where $(\theta_m',\phi_m')$ are the perturbed Bloch-sphere coordinates of the $m$th state, with corresponding subsystem states $Q^{S'}_m(Z,0)= \delta^Z_{\mathbf{Z}^{S'}_m}$.

Both the reference and perturbed states are evolved under the same Floquet operator and compared using the distinguishability measure $\Gamma$ and the state-space coverage index $\mathcal{S}_1$, which quantify finite-time instability and long-time exploration, respectively. The numerical parameters
used in these calculations are listed in Table~\ref{tab:numerical_parameters}
of Appendix~\ref{app:numerical_details}.
In the following subsections, we examine how varying the interaction strength $\kappa$ and system size $L$ drives a transition from periodic, confined dynamics to unstable and globally spreading behavior.

\begin{figure*}[htbp]
\centering

\begin{subfigure}{0.9\linewidth}
    \centering
    \hspace*{-1.2em}%
    \includegraphics[width=\linewidth]{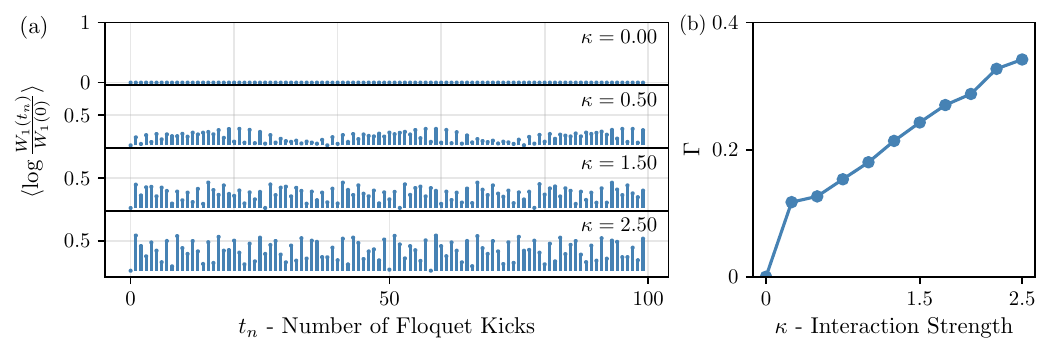}
\end{subfigure}

\vspace{0.5em}

\begin{subfigure}{0.9\linewidth}
    \centering
    \includegraphics[width=\linewidth]{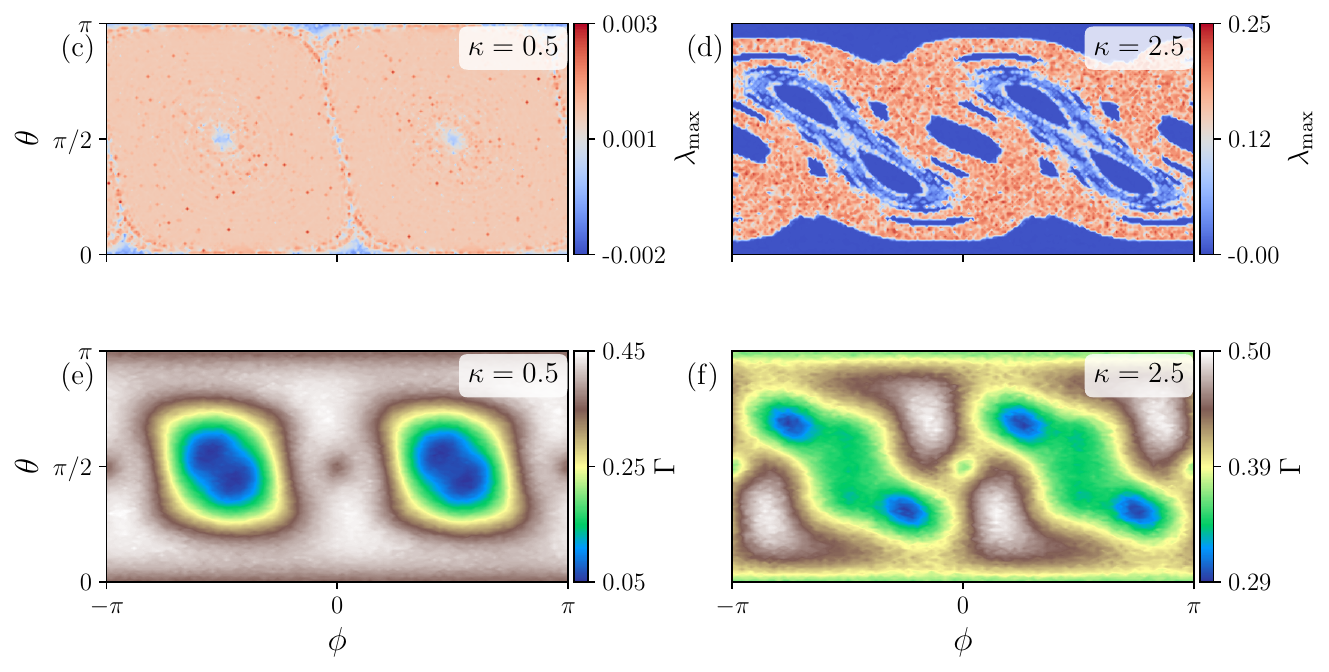}
\end{subfigure}

\caption{Distinguishability growth and phase-space structure in the three-qubit kicked top.
    (a) Time evolution of the average relative distance between perturbed and unperturbed GQSs for increasing interaction strength $\kappa$, starting from a spin-coherent state at $(\theta,\phi)=(\pi/2+0.5,\pi/2)$.
    (b) Corresponding distinguishability measure $\Gamma$ as a function of $\kappa$.
    (c,d) Classical maximal Lyapunov exponent $\lambda_{\max}$ over initial conditions $(\theta,\phi)$ for $\kappa=0.5$ and $2.5$.
    (e,f) Quantum distinguishability measure $\Gamma$ evaluated over spin-coherent product states for the same interaction strengths. Increasing $\kappa$ enhances both $\lambda_{\max}$ in the classical case and the distinguishability growth $\Gamma$ in the quantum case. The classical and quantum maps over initial conditions also exhibit qualitative similarities, indicating that the underlying kicked-top geometry continues to organize the reduced quantum dynamics.
    }
    \label{fig:F4}
\end{figure*}

\subsection{From Periodic Motion to Interaction-Induced Deformation}

We first examine how interactions reshape subsystem dynamics in the three-qubit kicked top, whose interaction graph reduces to a nearest-neighbor ring, allowing us to isolate the onset of interaction-induced complexity. Figure~\ref{fig:Int_Deform} shows the evolution of two nearby geometric quantum states of a qubit for increasing interaction strength $\kappa$, starting from a localized spin-coherent state.

For $\kappa = 0$, the subsystem state remains localized and follows a periodic trajectory on the Bloch sphere, confined to a narrow region of $\mathbb{C}P^1$, analogous to classical motion on invariant tori. Distinguishability between two nearby states is preserved at all times.
For $\kappa > 0$, this structure is progressively deformed. The distribution is sheared and stretched, probability mass redistributes across projective Hilbert space, and the motion loses periodicity as the geometric quantum state develops increasingly complex structure. At larger $\kappa$, this deformation leads to broad coverage of $\mathbb{C}P^1$. This transition mirrors the classical breakdown of invariant phase-space structures, now realized through the redistribution of probability mass on projective Hilbert space, and suggests interaction-induced sensitivity at the level of subsystem ensembles, which we quantify next.

\subsubsection{Local Instability: Distinguishability Measure}

To quantify the sensitivity suggested by Fig.~\ref{fig:Int_Deform}, we compute the distinguishability measure $\Gamma$ (Sec.~\ref{section04}), which captures the average logarithmic growth of the separation between nearby geometric quantum states.

We first consider the spin-coherent initial state $(\theta, \phi) = (\pi/2+0.5,\pi/2)$ and track the Wasserstein distance between nearby subsystem states for varying $\kappa$, as shown in Fig.~\ref{fig:F4}(a). In the absence of interactions ($\kappa = 0$), the distance remains constant, reflecting periodic dynamics. As $\kappa$ increases, the distance grows more strongly over time, indicating increasing sensitivity to perturbations.
This behavior is summarized in Fig.~\ref{fig:F4}(b), which shows $\Gamma$ as a function of $\kappa$. For small $\kappa$, $\Gamma$ remains close to zero, consistent with periodic or quasiperiodic dynamics, and it grows as $\kappa$ increases, signaling the onset of instability in subsystem dynamics.

To probe initial-state dependence, we compute $\Gamma$ across all spin-coherent states on the Bloch sphere. Figures~\ref{fig:F4}(e,f) show that $\Gamma$ depends strongly on the initial condition. For moderate $\kappa$, regions of low and high sensitivity coexist, while for larger $\kappa$, regions of enhanced instability expand across the state space. This structure closely parallels the phase-space behavior of the classical kicked top shown in Figs.~\ref{fig:F4}(c,d), where periodic and chaotic regions coexist, but here it emerges through the evolution of probability measures on projective Hilbert space.

\begin{figure*}[htbp]
    \centering

    \begin{subfigure}{0.95\linewidth}
        \centering
        \includegraphics[width=1.02\linewidth]{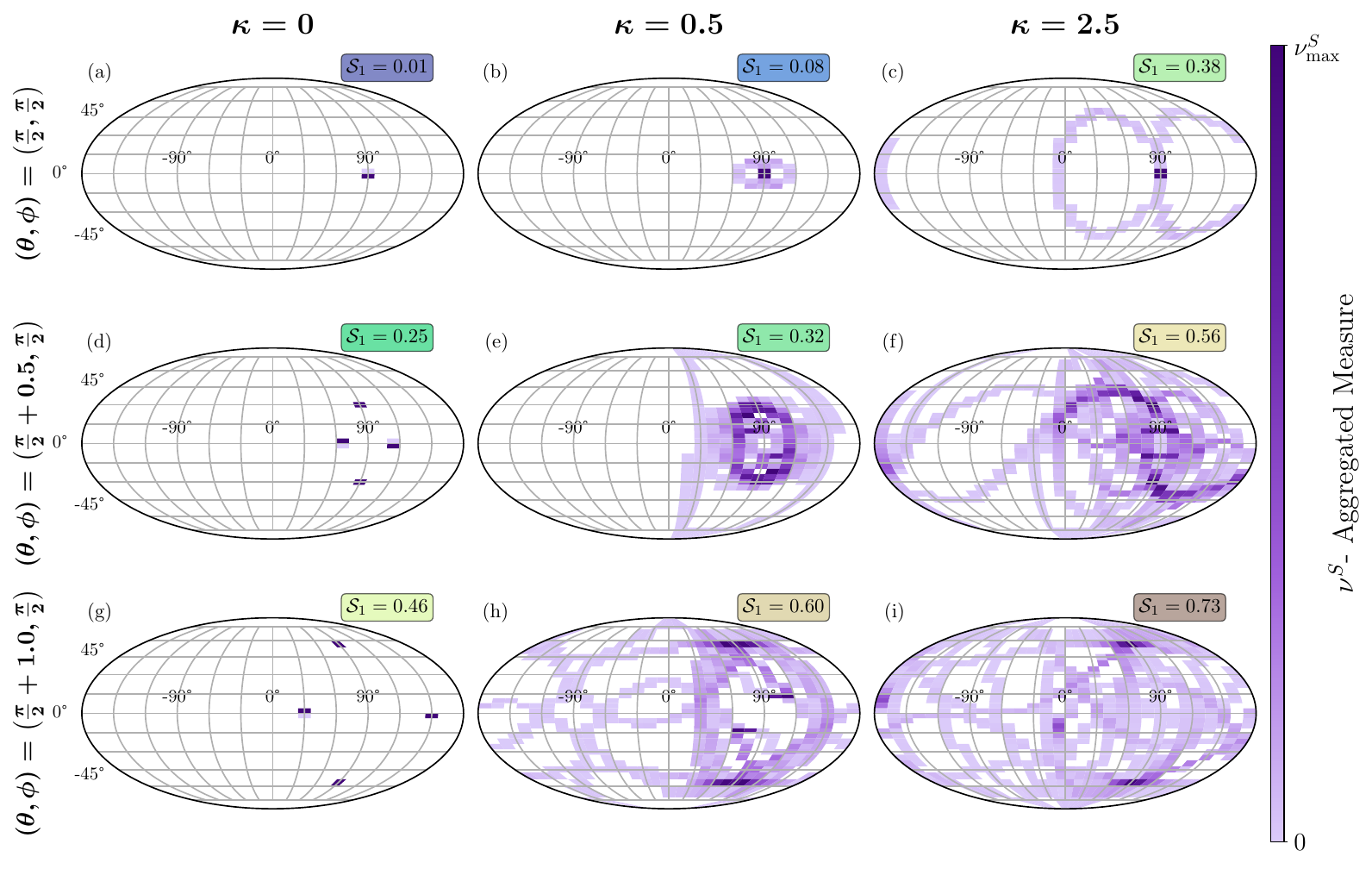}
    \end{subfigure}
    
    \begin{subfigure}{0.95\linewidth}
        \centering
        \includegraphics[width=1.0\linewidth]{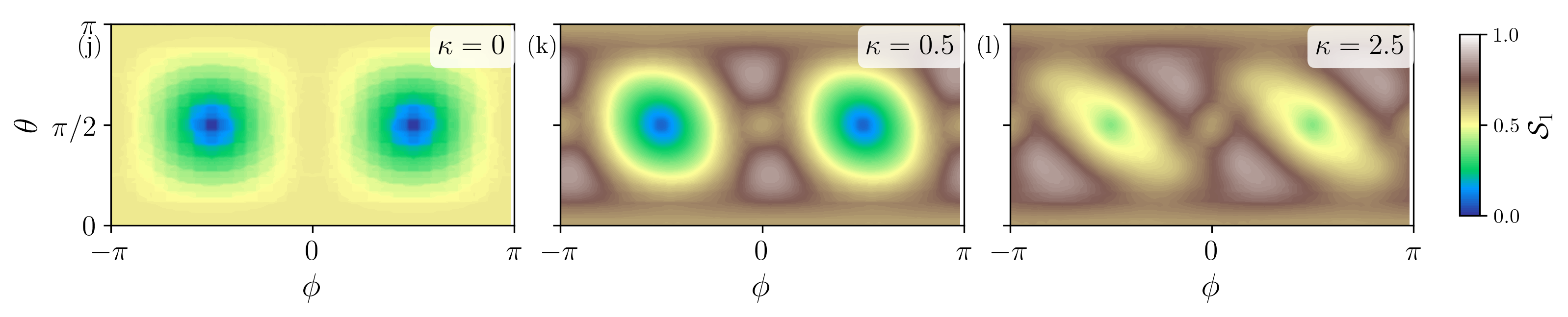}
    \end{subfigure}
    \caption{
    Interaction-induced spreading in the three-qubit kicked top.
    (a--i) Time-aggregated measures $\nu^S$ for three initial states (rows) and $\kappa=0, 0.5, 2.5$ (columns).
    For $\kappa=0$, dynamics remain periodic on $\mathbb{C}P^1$, yielding values of $\mathcal{S}_1$ set by orbit geometry. States near $(\pi/2,\pi/2)$ remain confined, while others explore larger regions.
    With increasing $\kappa$, probability mass is redistributed across the sphere, transitioning from structured spreading to broad coverage.
    (j--l) Corresponding $\mathcal{S}_1$ values over all initial states, showing increased exploration and persistent dependence on initial condition.
    }
    \label{fig:SSCI_combined}
\end{figure*}

These results show that interactions induce a transition from stable to unstable subsystem dynamics, with sensitivity that is both interaction dependent and geometrically organized across state space.

\subsubsection{Sensitivity and Spread}

To characterize long-time exploration, we use the State-Space Coverage Index $\mathcal{S}_1$ (Sec.~\ref{section04}). Figure~\ref{fig:SSCI_combined} shows the time-aggregated geometric quantum state for three representative initial conditions and increasing $\kappa$. For $\kappa=0$, the dynamics is periodic and $\Gamma=0$ for all initial states, yet $\mathcal{S}_1$ is generally nonzero, with a value set by the geometry of the periodic orbit on $\mathbb{C}P^1$. The spin-coherent state at $(\pi/2,\pi/2)$, an eigenstate of $J_y$, remains confined and yields small $\mathcal{S}_1$, while states farther away trace larger orbits and produce greater coverage.

As interactions are introduced ($\kappa>0$), probability mass is progressively redistributed across the Bloch sphere and $\mathcal{S}_1$ increases. For intermediate $\kappa$, the distribution spreads along structured regions, while for larger $\kappa$ it covers a substantial portion of $\mathbb{C}P^1$.

Figure~\ref{fig:SSCI_combined}(j--l) shows $\mathcal{S}_1$ over all spin-coherent states. As $\kappa$ increases, regions of low coverage shrink while regions of high coverage expand, reflecting a transition from confined motion to broad exploration. The structure of $\mathcal{S}_1$ over initial states mirrors that of $\Gamma$, with low sensitivity corresponding to low coverage and high sensitivity to high coverage, so the organization of instability is reflected in the organization of long-time spreading. Throughout, states farther from $(\pi/2,\pi/2)$ exhibit larger coverage, indicating that interactions enhance exploration without fully erasing the imprint of the initial state.

Together, $\Gamma$ and $\mathcal{S}_1$ show that dynamical complexity in the projected ensemble is governed not only by interaction strength but also by the location of the initial state on the Bloch sphere.

\subsection{From Interaction-Induced Deformation to Environment-Driven Complexity}

We now examine how environment size influences subsystem dynamics, considering systems of varying total size $L$ with fixed subsystem size $L_S = 1$ and environment size $L_E = L - 1$. For $L>3$, the non-nearest-neighbor couplings introduced by the collective $J_z^2$ term increase the redistribution of information across the system, enhancing entanglement generation and providing more channels through which perturbations can propagate. Both local instability ($\Gamma$) and global spreading ($\mathcal{S}_1$) are therefore affected.

\begin{figure*}
    \centering
    \includegraphics[width=1.\linewidth]{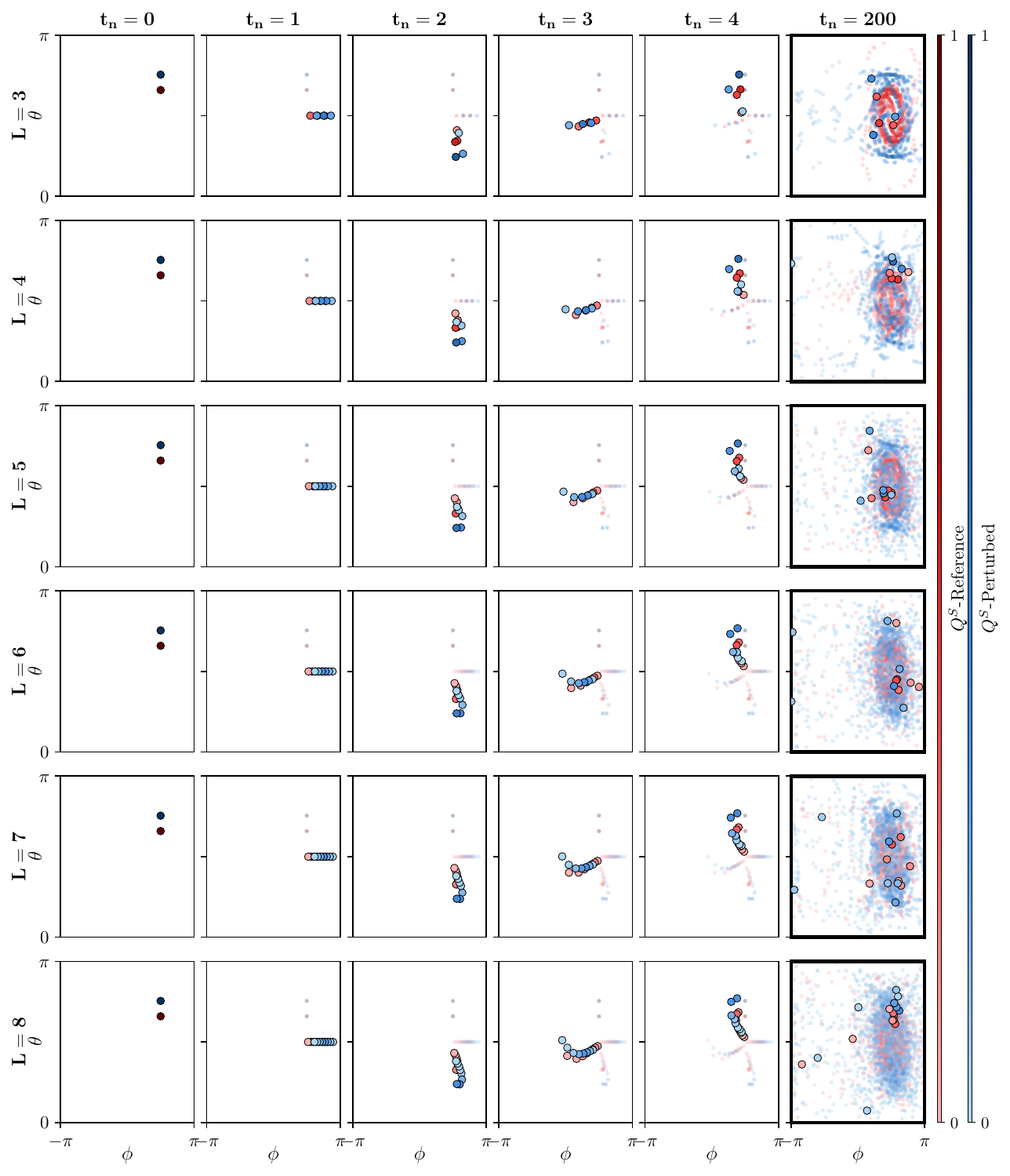}
    \caption{
    Evolution of two neighboring GQSs in the weakly interacting $L$-qubit quantum kicked top, $\kappa = 0.5$. Rows correspond to increasing total qubit number $L$, and columns show the states after Floquet kicks $t_n = 0, 1, 2, 3, 4, 200$. As $L$ increases, the ensembles exhibit greater sensitivity to initial perturbations and more pronounced deformation, while remaining confined to a limited region of $\mathbb{C}P^{1}$.
    }
    \label{fig:Env_Weak_Sensitivity}
\end{figure*}

\begin{figure*}
    \centering
    \includegraphics[width=1.\linewidth]{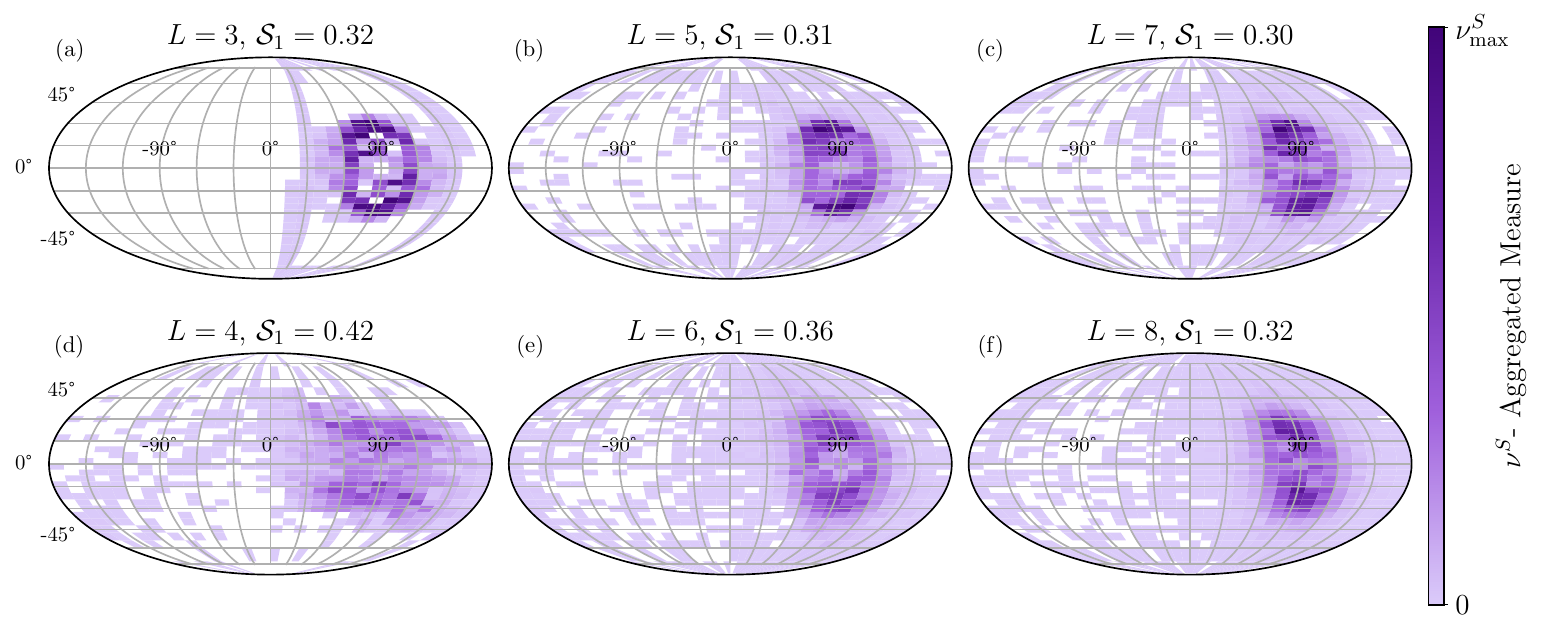}
    \caption{
    Time-aggregated GQSs $\nu^S$ and coverage index $\mathcal{S}_1$ in the weakly interacting quantum kicked top, $\kappa=0.5$. Top row (a--c) shows odd-$L$ systems, and bottom row (d--f) shows even-$L$ systems, with $L$ increasing across each row. The color scale gives the time-aggregated probability density on $\mathbb{C}P^1$. Even-$L$ systems exhibit slightly larger coverage than odd-$L$ systems, indicating weak finite-size parity-symmetry effects.
    }
    \label{fig:Env_Weak}
\end{figure*}

System size also determines the symmetry properties of the Hamiltonian. Even- and odd-qubit systems exhibit distinct parity symmetries, inherited by the Floquet operator, leading to qualitative differences in the dynamics \cite{paritydogra2019quantum,parityullah2026harnessing,RecurrenceSanthanam, RecurrencesGhose}. More generally, symmetries are known to constrain projected ensembles \cite{PEvarikuti2024unraveling,PEchang2025deep,PEyan2026characterizing}. In the present model, both the Floquet operator and the initial spin-coherent product state are invariant under permutations of the qubits. Conditional states associated with environment bit strings of equal Hamming weight therefore coincide, so $Q^S(Z,t)$ is supported on at most $L_E+1=L$ points of $\mathbb{C}P^1$ rather than $2^{L_E}$. In the following, we quantify how environment size shapes subsystem dynamics using $\Gamma$ and $\mathcal{S}_1$.

\subsubsection{Weak-Interaction Regime}

We first consider the weak-interaction regime, $\kappa = 0.5$, in which the dynamics retain significant structure, allowing us to isolate how environment size modifies subsystem behavior before the onset of strong mixing. Figure~\ref{fig:Env_Weak_Sensitivity} shows the evolution of two nearby subsystem states as $L$ increases. For small $L$, the motion is structured. As $L$ increases, the ensembles become more diffuse but retain visible organization, indicating that interactions are not yet strong enough to fully disrupt the underlying structure.

The corresponding time-aggregated states and coverage index $\mathcal{S}_1$, shown in Fig.~\ref{fig:Env_Weak}, remain localized and nonuniform, indicating limited state-space exploration. As $L$ changes, the distributions broaden and deform modestly, but the coverage remains constrained and does not grow monotonically. A weak parity dependence is already visible, with even-$L$ systems exhibiting slightly larger coverage than odd-$L$ systems. This regime provides a baseline in which environment effects are present but not yet dominant.

\subsubsection{Strong-Interaction Regime}

\begin{figure*}[htbp]
    \centering
    \includegraphics[width=1.\linewidth]{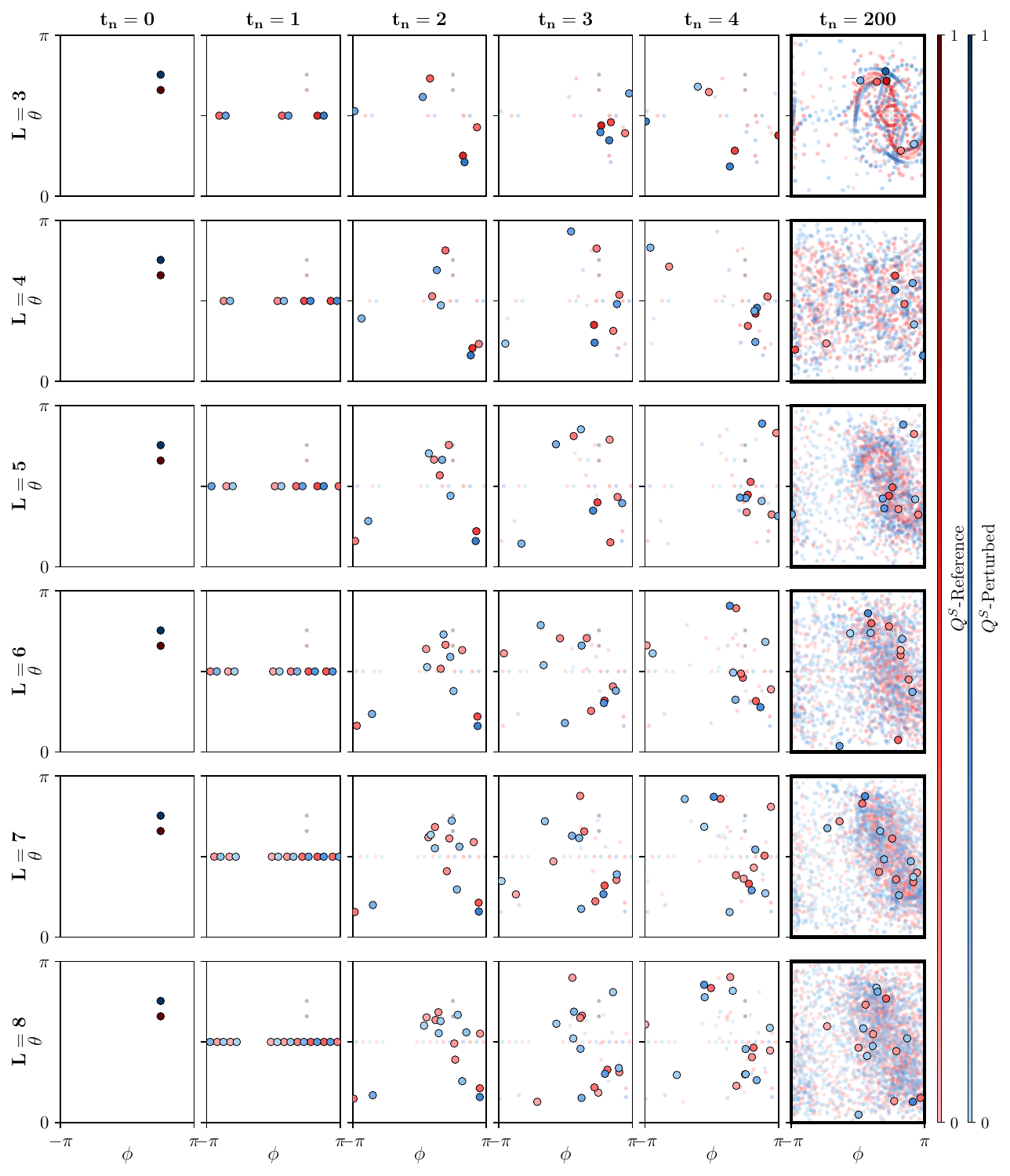}
    \caption{Evolution of two neighboring GQSs in the strongly interacting $L$-qubit quantum kicked top with $\kappa = 2.5$. Rows correspond to increasing values of $L$, while columns represent kicks at $t_n = 0, 1, 2, 3, 4, 200$. As $L$ increases, the ensembles spread across $\mathbb{C}P^{1}$, rapidly lose visible structure, and display greater sensitivity compared to the weakly interacting case. Systems with even $L$ tend to exhibit stronger sensitivity and more uniform spreading than those with odd $L$, highlighting finite-size parity-symmetry effects.}
    \label{fig:Env_Strong_Sensitivity}
\end{figure*}

\begin{figure*}[htbp]
    \centering
    \includegraphics[width=1.\linewidth]{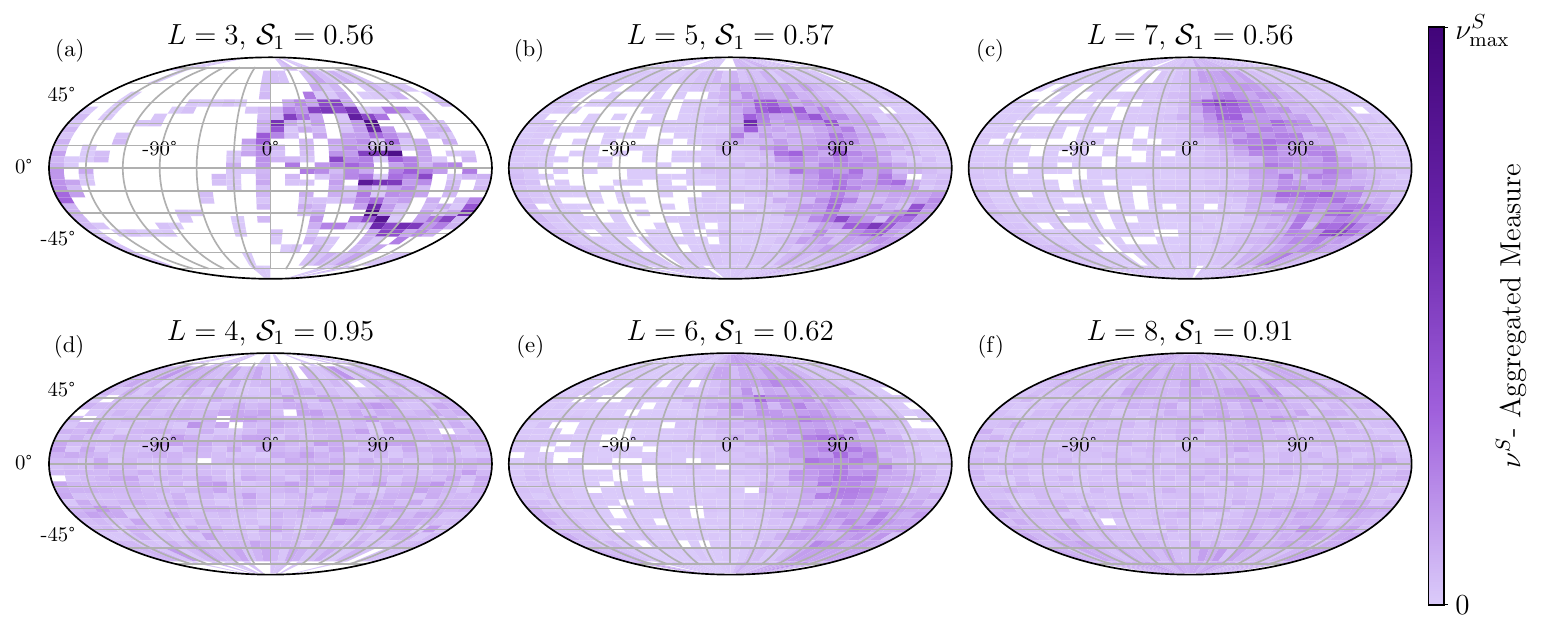}
    \caption{
    Time-aggregated GQSs $\nu^S$ and coverage index $\mathcal{S}_1$ in the strongly interacting quantum kicked top, $\kappa=2.5$. Top row (a--c) shows odd-$L$ systems, and bottom row (d--f) shows even-$L$ systems, with $L$ increasing across each row. The color scale gives the time-aggregated probability density on $\mathbb{C}P^1$. Even-$L$ systems exhibit larger coverage and more uniform spreading than odd-$L$ systems, revealing pronounced finite-size parity-symmetry effects.
    }
    \label{fig:Env_Strong}
\end{figure*}

In the strong-interaction regime, $\kappa = 2.5$, the dynamics exhibit substantial mixing and the influence of environment size becomes much more pronounced. Figure~\ref{fig:Env_Strong_Sensitivity} shows that the dynamics already display substantial deformation for small $L$. As $L$ increases, the ensembles rapidly lose structure and spread across the Bloch sphere, showing strong sensitivity and efficient propagation of perturbations.

The corresponding time-aggregated geometric quantum states and coverage index $\mathcal{S}_1$, shown in Fig.~\ref{fig:Env_Strong}, now cover a large fraction of $\mathbb{C}P^1$, approaching near-uniform coverage for larger even-$L$ systems and reflecting a transition to global exploration of state space. Parity effects are also more pronounced. Even-$L$ systems exhibit more uniform coverage and larger $\mathcal{S}_1$, while odd-$L$ systems retain residual structure, indicating that symmetry continues to organize the dynamics even under strong interactions. In this regime the dynamics are no longer constrained by underlying structure, and environment size plays a dominant role in driving complexity.

\subsubsection{Joint Dependence on Interaction Strength and Environment Size}

\begin{figure*}
    \centering
    \includegraphics[width=0.95\linewidth]{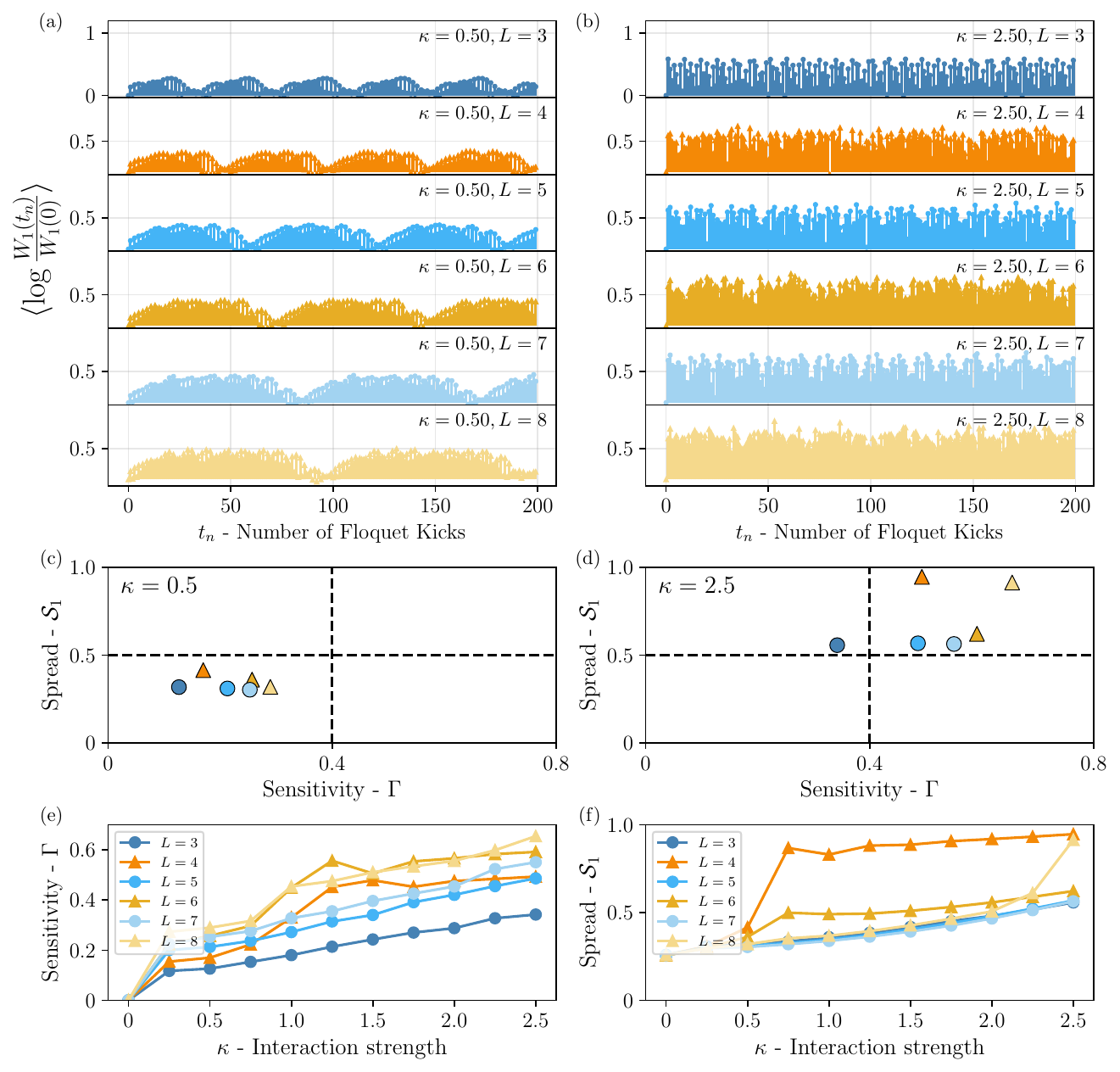}
    \caption{ Sensitivity and spreading in the $L$-qubit kicked top for the spin-coherent initial state $(\theta,\phi)=(\pi/2+0.5,\pi/2)$. (a,b) Separation dynamics of nearby subsystem states for different $L$ at $\kappa=0.5$ and $2.5$, showing weaker and stronger distinguishability growth, respectively. (c,d) Joint behavior of the distinguishability measure $\Gamma$ and the state-space coverage index $\mathcal{S}_1$. Dashed lines serve as visual guides to the qualitative dynamical regimes. (e,f) Dependence of $\Gamma$ and $\mathcal{S}_1$ on $\kappa$ for different $L$. Both generally increase with $\kappa$, while their dependence on $L$ is structured by parity symmetry, with even-$L$ systems often exhibiting larger values than neighboring odd-$L$ systems. }
    \label{fig:Env_Scaling}
\end{figure*}

We now examine the joint dependence of sensitivity $\Gamma$ and spread
$\mathcal{S}_1$ on interaction strength $\kappa$ and environment size $L$. The
separation dynamics in Figs.~\ref{fig:Env_Scaling}(a,b) provide a first
indication. Increasing $L$ increases both the magnitude and the fluctuations of
distinguishability growth, modestly for weak interactions and strongly for
strong interactions, showing that larger environments propagate perturbations
more effectively.

Placing each system in the $(\Gamma,\mathcal{S}_1)$ plane, as in
Figs.~\ref{fig:Env_Scaling}(c,d), makes this two-dimensional structure
explicit. The dashed lines indicate qualitative tendencies, following the
schematic of Fig.~\ref{fig:gamma_sp_regime_diagram}, rather than sharp
boundaries. At weak interaction, increasing $L$ raises sensitivity while
coverage remains limited, shifting the dynamics toward greater sensitivity
without a corresponding increase in exploration. Sensitivity and spread
therefore carry distinct information, since an environment can make subsystem
dynamics more unstable without making it more exploratory. At strong
interaction, both diagnostics are large, placing the dynamics toward the
sensitive, broadly exploring part of the plane. Within each parity sector, both
$\Gamma$ and $\mathcal{S}_1$ tend to increase with $L$, while the alternation
between even and odd $L$ makes their overall dependence on $L$ nonmonotonic.

Interaction strength is what moves systems across this plane. As
Figs.~\ref{fig:Env_Scaling}(e,f) show, both $\Gamma$ and $\mathcal{S}_1$
generally increase with $\kappa$ for every system size, so stronger coupling
promotes sensitivity and exploration together. Environment size sets how far
this progression reaches. The most pronounced change occurs between $L=3$ and
$L>3$, where the additional non-nearest-neighbor couplings substantially enrich
the dynamics.

Parity symmetry organizes this landscape. Even-$L$ systems generally lie
farther along both axes than neighboring odd-$L$ systems, and the separation
grows with $\kappa$ and $L$, where mixing is strongest. Symmetry thus
constrains not only how fast nearby projected ensembles separate but also how
much of state space they eventually cover. This parity-dependent organization
is a finite-size quantum effect and does not persist in the classical limit
$j \to \infty$ \cite{Recurrencehaake1987classical, RecurrencesGhose}.

These results show that the dynamical complexity of the projected ensemble is
two-dimensional and jointly controlled by interaction strength and system size. The interaction
strength $\kappa$ drives the transition from periodic to complex dynamics along
both axes, while the system size $L$, and hence the environment size $L-1$,
reshapes where systems sit in the $(\Gamma,\mathcal{S}_1)$ plane in a
parity-dependent and generally nonmonotonic manner.

\section{Conclusion and Outlook}
\label{section06}

In this work, we introduced a geometric framework for quantifying the dynamical
complexity of finite interacting quantum systems. Partitioning the system into
a subsystem and an environment, we represented the reduced dynamics by the
geometric quantum state, which coincides with the projected ensemble of the
many-body literature. Rather than asking how this ensemble approaches the
Haar distribution, we asked how sensitively it responds to perturbations of the
initial state and how extensively it explores state space over time. In this
picture, interactions turn the periodic motion of a localized pure state into
the deformation and spreading of a probability measure on
$\mathbb{C}P^{d_S-1}$.

To quantify these effects, we introduced two diagnostics motivated by classical
dynamics. The distinguishability measure $\Gamma$ is based on the Wasserstein
distance between nearby projected ensembles, and the state-space coverage index
$\mathcal{S}_1$ on the Wasserstein proximity of the time-aggregated ensemble to
the Haar measure. In the quantum kicked top, both generally increase with the
interaction strength $\kappa$, which drives the transition from periodic to
complex subsystem dynamics. Environment size acts more subtly. Within each
parity sector both diagnostics tend to increase with $L$, while the alternation
between even and odd $L$ makes their dependence on $L$ nonmonotonic. The two
diagnostics are not redundant. At weak interaction, larger environments
increase sensitivity without a corresponding significant increase in exploration, so
sensitivity and spread carry distinct information about the dynamics.
Permutation symmetry further constrains the projected ensemble,
restricting its support to at most $L$ points on $\mathbb{C}P^1$.

These diagnostics complement existing approaches to projected ensembles.
Studies of deep thermalization compare the instantaneous ensemble with the Haar
measure through its moments \cite{PEcotler2023emergent,PEippoliti2023dynamical},
whereas $\Gamma$ tracks how nearby ensembles separate in time and
$\mathcal{S}_1$ characterizes the time-aggregated ensemble. Because the
Wasserstein distance compares the ensembles themselves rather than their first
moments, it resolves rearrangements of probability mass that leave the reduced
density matrix unchanged and are therefore invisible to operator-level
distances such as the Bures distance. The time-aggregated ensemble also
connects to Hilbert-space ergodicity
\cite{PEmark2024maximum,PEpilatowsky2024hilbert}, with environment conditioning
allowing broad subsystem coverage even when periodic driving obstructs global
ergodicity. More broadly, the approach is state-based rather than
operator-based, complementing out-of-time-ordered correlators and Loschmidt
echoes as well as Husimi-based phase-space measures \cite{EEtang2026phase}, and
it applies at the finite, experimentally relevant system sizes considered here.

Several directions follow directly from the present results. All diagnostics
here were computed by conditioning in the computational basis, and since the
measurement basis is known to shape projected ensembles
\cite{PEippoliti2023dynamical,PEvarikuti2024unraveling,PEliu2026coherence},
mapping how $\Gamma$ and $\mathcal{S}_1$ change across bases is a natural next
step. Scaling to larger systems is aided by the symmetry identified above.
Because permutation invariance confines the dynamics to the symmetric subspace
and restricts the projected ensemble to at most $L$ points, much larger $L$ and
subsystems with $L_S>1$ are accessible for the kicked top, while tensor-network
methods offer a route to models without this symmetry. The qualitative agreement between $\Gamma$ and the classical maximal Lyapunov
exponent in Fig.~\ref{fig:F4} also motivates a quantitative comparison in the
classical limit $j\to\infty$, together with a systematic comparison with
operator-based diagnostics of quantum chaos such as OTOCs \cite{OTOCLEchoQKT}.

On the experimental side, the kicked top admits a direct circuit implementation
as alternating collective rotations and $ZZ$ interactions, and projected
ensembles are already accessible on current quantum simulators
\cite{PEchoi2023preparing,PEyan2026characterizing}. Measuring $\Gamma$ and
$\mathcal{S}_1$ on hardware, which is the subject of ongoing work, raises the
question of noise. Under noise the conditional states become mixed
\cite{PEyan2026characterizing}, which calls for extending the transport
framework from $\mathbb{C}P^{d_S-1}$ to the full state space and makes the
diagnostics a potential probe of how hardware noise reshapes subsystem
dynamics.

These results establish a geometric, state-resolved perspective on quantum
dynamical complexity in finite interacting systems, where interaction-induced deformation and spreading of projected ensembles
capture instability and exploration in a manner inspired by classical dynamics.

\section*{Acknowledgements} \label{sec:acknowledgements}
The authors thank Jinghao Lyu and Gregory Wimsatt for helpful discussions. Code and data supporting this work are available at \url{https://github.com/Kommmi/Qaos}. This material is based on work supported by, or in part by, the MELICERTES project (ANR-22-PEAE-0010) of the French National Research Agency (France2030, national PEPR “agroécologie et numérique” programmes), Inria’s CONCAUST Exploratory Action, Templeton World Charity Foundation grant TWCF0570, the U.S. Army Research Laboratory and U.S. Army Research Office grant W911NF-21-1-0048, and by the Art and Science Laboratory via a gift to UC Davis' Complexity Sciences Center.

\bibliographystyle{unsrt}
\bibliography{ref}  

\appendix*
\appendix

\section*{Appendices}
\noindent
The appendices provide additional conceptual, computational, and numerical
details supporting the main text. Appendix~\ref{app:GQS_vs_Schmidt} compares the geometric quantum state (GQS), or projected ensemble, representation with the Schmidt decomposition,
while Appendix~\ref{app:GQS_vs_Husimi} distinguishes GQSs from Husimi representations.
Appendix~\ref{app:wasserstein} describes the computation of the Wasserstein distance on
$\mathbb{CP}^{d_S-1}$, and Appendix~\ref{app:information_dimension} relates the State-Space Coverage Index
to the classical information dimension. Appendix~\ref{app:numerical_details} summarizes the numerical
implementation and simulation parameters. Appendix~\ref{app:quantum_recurrence} examines quantum
recurrence and its relation to distinguishability growth and state-space
coverage. Finally, Appendix~\ref{app:phase_space_structure} explores how these diagnostics vary across the
spin-coherent-state phase space of the $L$-qubit kicked top.

\section{Geometric Quantum States and Schmidt Decomposition}
\label{app:GQS_vs_Schmidt}
For a pure bipartite state $|\Psi_{SE}(t)\rangle \in \mathcal{H}_S \otimes \mathcal{H}_E$, the Schmidt decomposition is
\begin{equation}
|\Psi_{SE}(t)\rangle
=
\sum_{\alpha=1}^{r}
\sqrt{\lambda_\alpha(t)}\,
|\tilde{s}_\alpha(t)\rangle \otimes |\tilde{e}_\alpha(t)\rangle,
\label{eq:schmidt_decomp}
\end{equation}
where $\lambda_\alpha(t)\ge 0$, $\sum_\alpha \lambda_\alpha(t)=1$, and $r \le \min(\dim\mathcal{H}_S,\dim\mathcal{H}_E)$. Tracing over the environment gives
\begin{equation}
\rho_S(t)
=
\sum_{\alpha=1}^{r}
\lambda_\alpha(t)\,
|\tilde{s}_\alpha(t)\rangle\langle \tilde{s}_\alpha(t)|.
\label{eq:schmidt_spectral_rho}
\end{equation}
The quantities $\lambda_\alpha$ are the eigenvalues of $\rho_S(t)$,
while $\sqrt{\lambda_\alpha}$ are the Schmidt coefficients.

In the geometric quantum framework, instead of diagonalizing $\rho_S(t)$, we consider an environment-conditioned decomposition with respect to a fixed environment basis $\{|e_j\rangle\}$,
\begin{equation}
|\Psi_{SE}(t)\rangle
=
\sum_{j=1}^{d_E}
\sqrt{\lambda_j^{E}(t)}\,
|\chi_j^{S}(t)\rangle\otimes|e_j\rangle,
\label{eq:env_resolved_purestate}
\end{equation}
where
\begin{equation}
\lambda_j^{E}(t) = \sum_{k=1}^{d_S}|\psi_{kj}(t)|^2,
\quad
|\chi_j^{S}(t)\rangle
=
\frac{1}{\sqrt{\lambda_j^{E}(t)}}
\sum_{k=1}^{d_S}\psi_{kj}(t)\,|s_k\rangle.
\end{equation}
Tracing over the environment then gives
\begin{equation}
\rho_S(t)
=
\sum_{j=1}^{d_E}
\lambda_j^{E}(t)\,
|\chi_j^{S}(t)\rangle\langle \chi_j^{S}(t)|.
\label{eq:env_resolved_convex_rho}
\end{equation}

Unlike the Schmidt decomposition, the states $\{|\chi_j^{S}(t)\rangle\}$ are not generally orthogonal, and the weights $\{\lambda_j^{E}(t)\}$ are not eigenvalues of $\rho_S(t)$. Within the GQS framework, the pairs $\{\lambda_j^{E}(t),|\chi_j^{S}(t)\rangle\}$ define a probability distribution over subsystem pure states, which is the projected ensemble associated with the basis $\{|e_j\rangle\}$. This representation preserves the explicit correspondence between environment configurations and subsystem states. This association with the chosen environment basis is lost after rotation to the Schmidt basis.

Thus, while the Schmidt decomposition characterizes the magnitude of entanglement through its spectrum, the environment-resolved decomposition provides the structural information needed to study the geometry and transport of subsystem state distributions.

\section{Geometric Quantum States and Husimi Representations}
\label{app:GQS_vs_Husimi}

\begin{figure}[ht]
\centering
\begin{tikzpicture}[
  node distance=1.3cm,
  box/.style={draw, rounded corners, inner sep=4pt, align=center},
  arrow/.style={-Latex, thick}
]

\node[box] (cp) {$\mathbb{C}P^{d_S-1}$\\[-2pt]\footnotesize (pure state manifold)};

\node[box, below=of cp] (P) {$\mathcal{P}\!\left(\mathbb{C}P^{d_S-1}\right)$\\[-2pt]\footnotesize (GQS / probability measures)};

\node[box, below=of P] (D) {$\mathcal{D}_{d_S}$\\[-2pt]\footnotesize (density matrices)};

\node[box, below=of D] (F) {$\mathcal{F}\!\left(\mathbb{C}P^{d_S-1}\right)$\\[-2pt]\footnotesize (functions / densities on projective space)};

\draw[arrow] (cp) -- node[right] {\footnotesize \shortstack{environmental\\interactions}} (P);
\draw[arrow] (P) -- node[right] {\footnotesize first moment} (D);
\draw[arrow] (D) -- node[right] {\footnotesize coherent-state projection} (F);

\node[below=6pt of F, align=center] {\footnotesize
$\displaystyle Q_H(\psi)=\langle \psi|\rho_S(t)|\psi\rangle,
\quad
Q_H:\mathbb{C}P^{d_S-1}\to\mathbb{R}_{\ge 0}$};

\end{tikzpicture}
\caption{
Geometric quantum states (GQS) describe subsystem states as probability measures on $\mathbb{C}P^{d_S-1}$. The reduced density matrix $\rho_S$ is obtained as the first moment of the GQS and can be used to construct a Husimi representation through projection onto coherent states. Unlike the GQS, which is a discrete measure not determined by $\rho_S$ alone, the Husimi representation is, up to normalization, a smooth probability density on $\mathbb{C}P^{d_S-1}$ fixed entirely by $\rho_S$.
}
\label{fig:gqs_pushforward_husimi}
\end{figure}
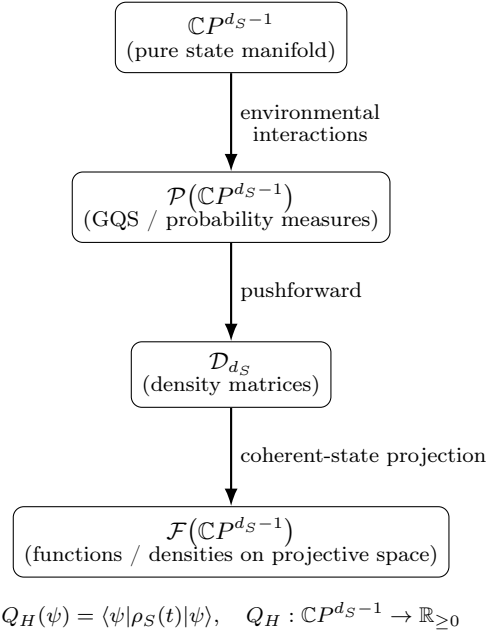

In quantum mechanics, mixed states are often visualized through phase-space representations such as the Husimi $Q$-function. In this work, we instead employ the geometric quantum state (GQS), which arises naturally from the system--environment decomposition and admits a direct geometric interpretation. The relationship between these representations is summarized schematically in Fig.~\ref{fig:gqs_pushforward_husimi}.

\begin{figure*}
    \centering
     \includegraphics[width=1.\linewidth]{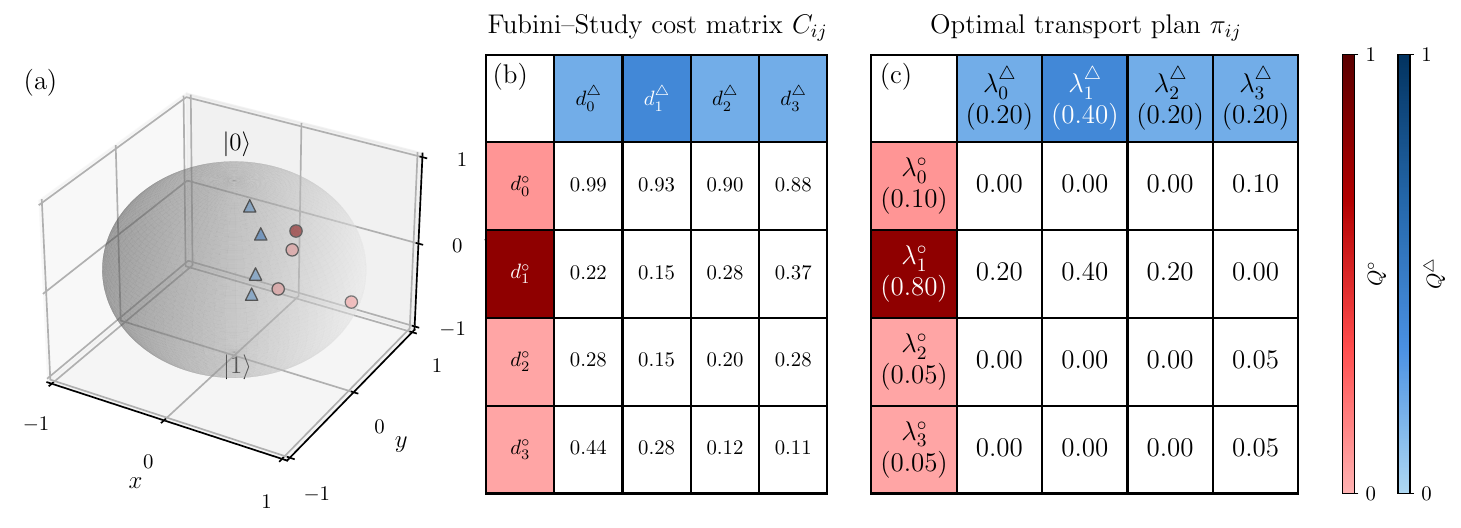}
    \caption{
    Geometric computation of the Wasserstein distance between two geometric quantum states (GQSs).
    (a) Discrete probability measures $Q^\Delta$ and $Q^\circ$ on the Bloch sphere $\mathbb{C}P^1$.
    (b) Fubini--Study distance cost matrix $C_{ij}$ between support points.
    (c) Optimal transport plan $\pi_{ij}$, illustrating how probability mass is redistributed to obtain $W_1(Q^\Delta, Q^\circ)=0.27$.
    }
    \label{fig:EMDExplain}
\end{figure*}

The GQS representation introduced in Sec.~\ref{section02} expresses subsystem states as probability measures on projective Hilbert space. For completeness, we recall that
\begin{equation}
Q^S(Z,t) = \sum_{j=1}^{d_E} \lambda_j^E(t)\, \delta^Z_ {Z_j^S(t)} \in \mathcal{P}(\mathbb{C}P^{d_S-1}),
\end{equation}
where the weights and conditional states are determined by the system--environment decomposition of the global wavefunction (see Sec.~\ref{section02}).

While the reduced density matrix $\rho_S(t)$ captures all observable statistics, it does not uniquely specify how probability is distributed across pure states. The GQS representation retains this ensemble structure explicitly, enabling one to track how probability mass is organized and evolves on $\mathbb{C}P^{d_S-1}$ under interactions.

Given a reduced density matrix $\rho_S(t)$, the Husimi function is defined as
\begin{equation}
Q_H(\psi) = \langle \psi | \rho_S(t) | \psi \rangle,
\end{equation}
where $|\psi\rangle \in \mathbb{C}P^{d_S-1}$ denotes a chosen family of coherent or reference states. This defines a nonnegative function on projective Hilbert space which, up to normalization, is a smooth probability density, and provides a visualization of observable statistics.

The key distinction between the GQS and Husimi representations lies in how they encode ensemble structure. Distinct probability distributions on $\mathbb{C}P^{d_S-1}$ may correspond to the same density matrix $\rho_S(t)$ and are therefore indistinguishable by any measurement on the subsystem. Since the Husimi function is constructed from $\rho_S(t)$, it inherits this indistinguishability: geometrically distinct GQS that share the same density matrix yield identical Husimi representations.

Consequently, while the Husimi representation is well-suited for visualizing observable statistics, the GQS framework preserves the underlying ensemble structure, enabling the study of dynamical complexity through the evolution and transport of probability mass on projective Hilbert space.

\section{Computation of Wasserstein Distance on \texorpdfstring{$\mathbb{C}P^{d_S-1}$}{CP(dS-1)}}
\label{app:wasserstein}
The definition of the Wasserstein distance on projective Hilbert space is given in Sec.~\ref{sec:section03}. Here, we summarize its formulation and numerical implementation for the geometric quantum states considered in this work.

For geometric quantum states of the form
\begin{equation}
Q = \sum_i \lambda_i\, \delta^Z_{Z_i},
\qquad
Q' = \sum_j \lambda'_j\, \delta^Z_{Z'_j},
\end{equation}
the computation of the Wasserstein distance reduces to a finite-dimensional optimal transport problem. The cost of transporting probability mass between points $Z_i$ and $Z'_j$ is given by the Fubini--Study distance,
\begin{equation}
C_{ij} = d_{FS}(Z_i,Z'_j).
\end{equation}

The Wasserstein distance $W_1(Q,Q')$ is then obtained by solving for the optimal transport plan $\pi_{ij}$ that minimizes the total transport cost,
\begin{equation}
W_1(Q,Q') = \min_{\pi\in\Pi(Q,Q')} \sum_{i,j} \pi_{ij}\, C_{ij},
\end{equation}
subject to the constraints
\begin{equation}
\pi_{ij}\ge 0,
\qquad
\sum_j \pi_{ij} = \lambda_i,
\qquad
\sum_i \pi_{ij} = \lambda'_j.
\end{equation}

This formulation corresponds to a linear programming problem over the transport plan $\pi_{ij}$. Figure~\ref{fig:EMDExplain} illustrates the procedure for two GQSs on the Bloch sphere, showing the support points, the Fubini--Study cost matrix, and the resulting optimal transport plan.

In practice, the computation proceeds by constructing the cost matrix $C_{ij}$ using pairwise Fubini--Study distances between the support points of the two geometric quantum states. For geometric quantum states with at most $d_E$ support points, the cost matrix has size at most $d_E \times d_E$, so the dominant computational cost comes from solving the discrete optimal transport problem over these environment-conditioned components rather than from the subsystem dimension $d_S$ itself. The optimal transport problem is then solved numerically to obtain the transport plan $\pi_{ij}$ and the corresponding Wasserstein distance.

In this work, we use the exact Earth Mover's Distance (EMD) solver of the Python Optimal Transport library \cite{OTNflamary2021pot}. For the system sizes considered here, the discrete formulation remains tractable and allows for direct evaluation of the Wasserstein distance without additional approximations.

This discrete optimal transport formulation enables a direct comparison of subsystem states by quantifying how probability mass is redistributed across projective Hilbert space.

\section{State-Space Coverage Index and Information Dimension}
\label{app:information_dimension}

In Sec.~\ref{section04}, we introduced the State-Space Coverage Index (SSCI) as a measure of long-time dynamical behavior, inspired by the classical notion of information dimension. Here, we briefly review the definition of information dimension and explain the challenges in extending it to the present setting.

In classical dynamical systems, long-time geometric complexity is often characterized through the information dimension of an invariant measure. Let $T : M \to M$ be a dynamical map on a compact metric space $(M,d)$, and let $\mu \in \mathcal{P}(M)$ be an invariant probability measure. $\mu$ can be approximated by the empirical (time-averaged) measure generated by a trajectory $\{x_n\}_{n=0}^{N-1}$,
\begin{equation}
\mu_N(A)
=
\frac{1}{N}
\sum_{n=0}^{N-1}
\mathbf{1}_A(x_n),
\qquad
A \subset M,
\label{eq:empirical_measure_appendix}
\end{equation}
which converges to $\mu$ under appropriate ergodicity assumptions.

To probe the geometric structure of $\mu$, one partitions $M$ into sets $\{\Delta_i(\epsilon)\}$ of diameter $\epsilon$, defining probabilities
\begin{equation}
p_i(\epsilon) = \mu(\Delta_i(\epsilon)).
\end{equation}
The associated Shannon entropy at scale $\epsilon$ is
\begin{equation}
H(\epsilon)
=
- \sum_i p_i(\epsilon)\,\log p_i(\epsilon),
\end{equation}
and the information dimension is defined as
\begin{equation}
D_1
=
\lim_{\epsilon \to 0}
\frac{H(\epsilon)}{\log(1/\epsilon)},
\label{eq:information_dimension_appendix}
\end{equation}
whenever the limit exists.

Geometrically, $D_1$ characterizes how probability mass is distributed across phase space at fine scales. For smooth measures, it coincides with the dimension of the underlying space, while for singular measures (e.g., strange attractors), it can take non-integer values, reflecting fractal structure.

In the geometric quantum framework considered here, the dynamical object is not a trajectory of points in $M = \mathbb{C}P^{d_S-1}$, but a trajectory of probability measures
\[
Q^{S}(\cdot,t) \in \mathcal{P}(M).
\]

A direct analogue of the empirical construction in Eq.~\eqref{eq:empirical_measure_appendix} would therefore yield a measure over measures,
\begin{equation}
\mathbb{M}_T
=
\frac{1}{T}
\sum_{t=0}^{T-1}
\delta_{Q^{S}(\cdot,t)}
\in
\mathcal{P}\!\left(\mathcal{P}(M)\right),
\end{equation}
which resides on an infinite-dimensional space. Constructing partitions and defining entropy scaling in this space is not straightforward and becomes computationally intractable upon discretization.

For this reason, we adopt the State-Space Coverage Index introduced in the main
text as a tractable alternative, providing a measure of how extensively
subsystem states spread across projective Hilbert space over long times.

\section{Numerical Implementation Details}
\label{app:numerical_details}
Numerical simulations are performed as described in the main text, including
the construction of geometric quantum states, computation of Wasserstein
distances via discrete optimal transport, and evaluation of the dynamical
diagnostics $\Gamma$ and the State-Space Coverage Index.
Table~\ref{tab:numerical_parameters} lists the numerical parameters used in
these calculations. A complete implementation of the methods used in this work
is available at \url{https://github.com/Kommmi/Qaos}.

\begin{table*}[t]
\centering
\caption{
Numerical parameters used to compute the distinguishability measure
$\Gamma$ and the state-space coverage index $\mathcal{S}_1$.
}
\label{tab:numerical_parameters}
\renewcommand{\arraystretch}{1.25}
\begin{tabular}{p{0.25\textwidth} p{0.30\textwidth} p{0.40\textwidth}}
\hline\hline
\textbf{Category} & \textbf{Parameter} & \textbf{Value or implementation detail} \\
\hline
Geometric quantum states
& Environment basis
& Computational basis \\
& Zero-weight cutoff
& Conditional states with $\lambda_j^E < 10^{-12}$ are omitted \\
& State representation
& Normalized state vectors / Bloch-sphere coordinates \\
\hline
Optimal transport
& Wasserstein order
& $p=1$ \\
& Ground metric
& Fubini--Study distance \\
& Optimal-transport solver
& EMD (POT) \\
\hline
Distinguishability measure $\Gamma$
& Time step
& $t_n=n$, with $\tau=1$ \\
& Number of perturbations
& $M = 200$ \\
& Perturbation scale
& Rotation angles drawn from $\alpha\sim\mathcal{N}(0,\epsilon^2)$, with
  $\epsilon=0.2$ radians \\
& Perturbation distribution
& Isotropic random rotations with uniformly distributed rotation axes
  and Gaussian-distributed rotation angles \\
\hline
Coverage index $\mathcal{S}_1$
& Number of bins over Bloch sphere
& $30 \times 30$ \\
& Binning method
& Equal-area bins on the Bloch sphere \\
& Averaging time
& $T_{\mathcal{S}} = 5000$ \\
\hline\hline
\end{tabular}
\end{table*}

\section{Complexity and Quantum Recurrence}
\label{app:quantum_recurrence}
\begin{figure*}
    \centering
    \begin{subfigure}{0.9\linewidth}
    \includegraphics[width=1\linewidth]{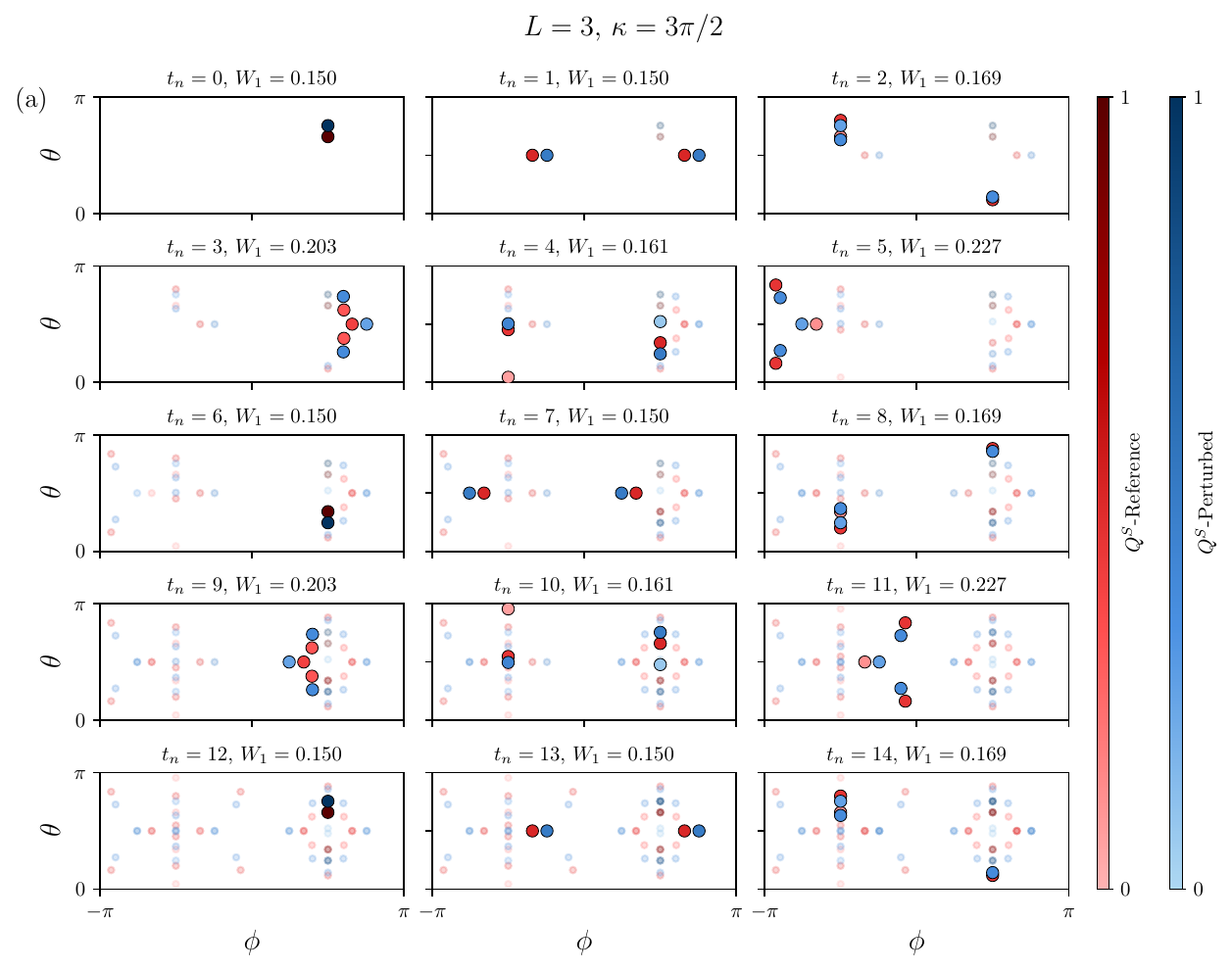}
    \end{subfigure}
    \hfill
    \begin{subfigure}{0.9\linewidth}
    \includegraphics[width=1\linewidth]{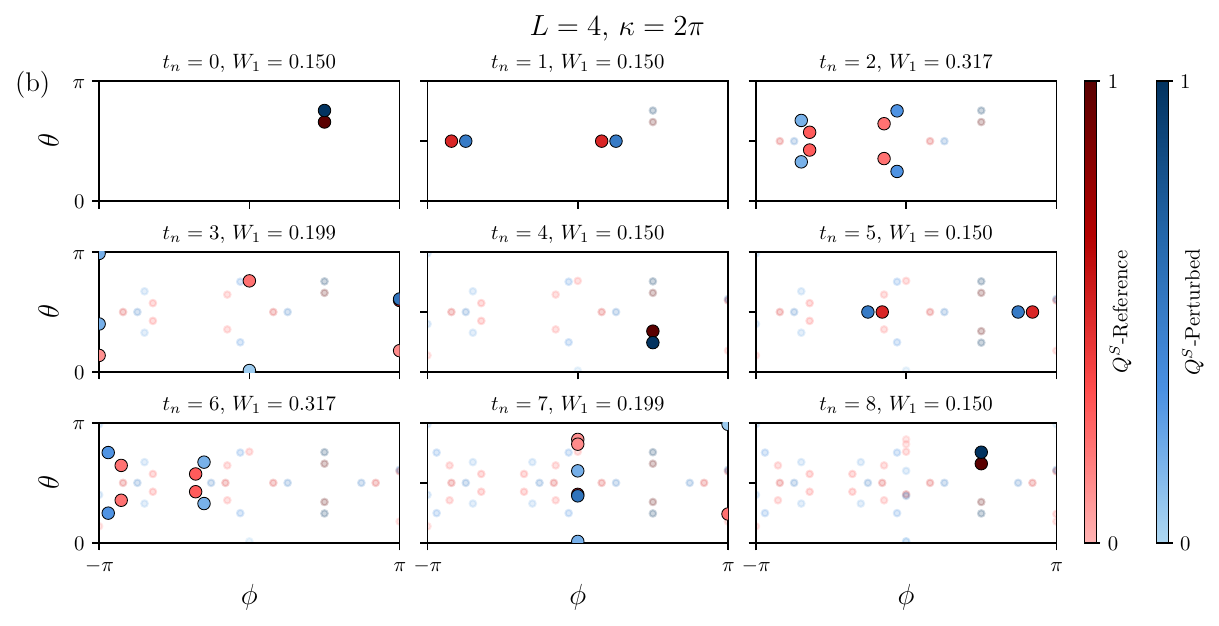}
    \end{subfigure}
    \caption{
Subsystem dynamics of two nearby GQSs and their Wasserstein separation in the recurrent regime \(\kappa=\pi j\). (a) \(L=3\), \(\kappa=3\pi/2\). (b) \(L=4\), \(\kappa=2\pi\). Recurrence occurs after 12 and 8 kicks, respectively. Although the dynamics are recurrent, distinguishability is not preserved for either the integer- or half-integer-spin kicked top at \(\kappa=\pi j\).
    }
    \label{fig:R1_t}
\end{figure*}

\begin{figure*}[htbp]
    \centering
    \begin{subfigure}{0.99\linewidth}
         \includegraphics[width=0.85\linewidth]{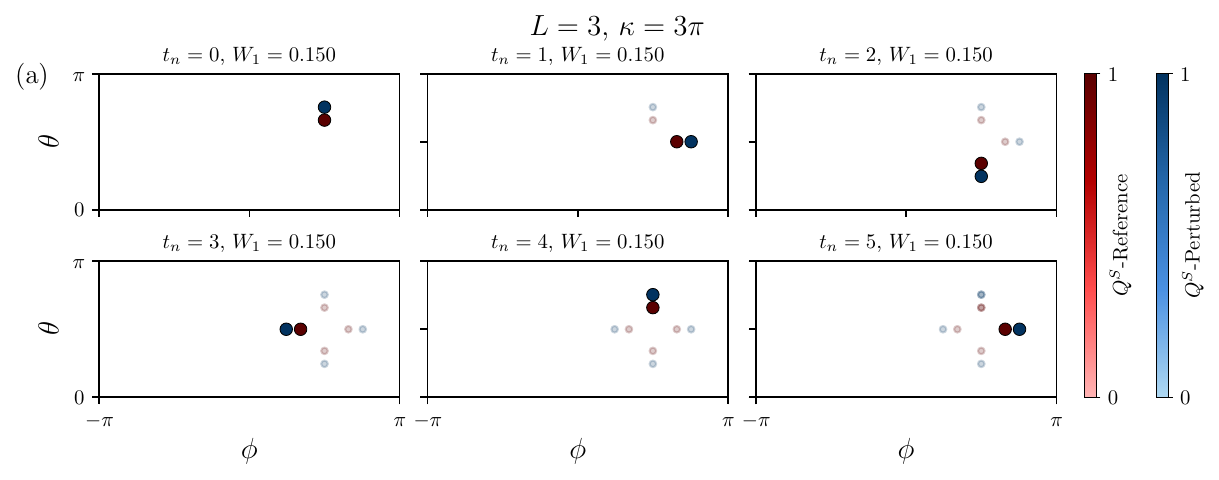}
    \end{subfigure}
    \hfill
    \begin{subfigure}{0.99\linewidth}
         \includegraphics[width=0.85\linewidth]{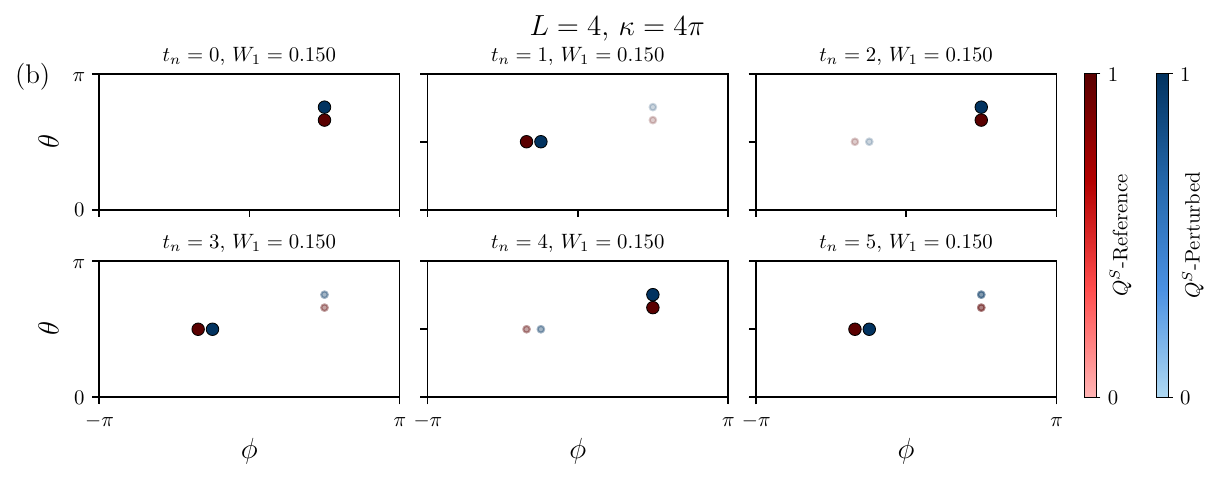}
    \end{subfigure}
    \caption{Subsystem dynamics of two nearby GQSs and their separation. (a) $(L,\kappa)=(3,3\pi)$ and (b) $(L,\kappa)=(4,4\pi)$. Recurrence occurs after 4 and 2 kicks, respectively, with separation preserved for $\kappa=2\pi j$.}
    \label{fig:R2}
\end{figure*}
In the main text, we examined complexity for $\kappa \in [0,2.5]$. Here, we focus on interaction strengths at which quantum recurrences are known to occur \cite{RecurrencesGhose, paritydogra2019quantum, RecurrenceSanthanam}. We consider the quantum kicked top for both half-integer spins $j \in \{3/2,5/2,7/2\}$ ($L=3,5,7$) and integer spins $j \in \{2,3,4\}$ ($L=4,6,8$), allowing us to probe behavior across different parity symmetries.

\subsection{Interaction strength \texorpdfstring{$\kappa = \pi j$}{kappa = pi j}}
We examine the evolution of two nearby GQSs initialized from spin-coherent states $(\theta,\phi) = (\pi/2+0.5,\pi/2)$ (circles) and $(\pi/2+0.8,\pi/2)$ (triangles). As shown in Fig.~\ref{fig:R1_t}, the subsystem exhibits recurrence after 12 kicks for $L=3$ (half-integer spin) and 8 kicks for $L=4$ (integer spin), consistent with theoretical predictions \cite{RecurrencesGhose}.

However, recurrence of the trajectory does not imply preservation of nearby-state separation. As shown in Fig.~\ref{fig:R1}(a), the distinguishability between the two GQSs undergoes oscillatory growth rather than remaining fixed. Thus, unlike the noninteracting case $\kappa=0$, the subsystem dynamics are recurrent but distinguishability is not preserved. This yields a small but positive distinguishability measure, $\Gamma>0$, indicating sensitivity to perturbations.

Figure~\ref{fig:R1}(c) summarizes this behavior in the sensitivity--spread plane, where $\Gamma$ quantifies average distinguishability growth and $\mathcal{S}_1$ quantifies long-time state-space coverage. As in Sec.~\ref{section05}, integer-spin systems, corresponding to even $L$, exhibit larger sensitivity and spread than half-integer-spin systems, corresponding to odd $L$. Relative to the schematic of Fig.~\ref{fig:gamma_sp_regime_diagram}, the $\kappa=\pi j$ recurrent dynamics therefore lie toward relatively large coverage but low sensitivity.

\subsection{Interaction strength \texorpdfstring{$\kappa = 2\pi j$}{kappa = 2pi j}}
We now consider the same initial states for interaction strength $\kappa = 2\pi j$. As shown in Fig.~\ref{fig:R2}, recurrence occurs after 4 kicks for $L=3$ and 2 kicks for $L=4$, again consistent with theoretical predictions \cite{RecurrencesGhose}.

\begin{figure*}[htbp]
    \centering
    \includegraphics[width=1.\linewidth]{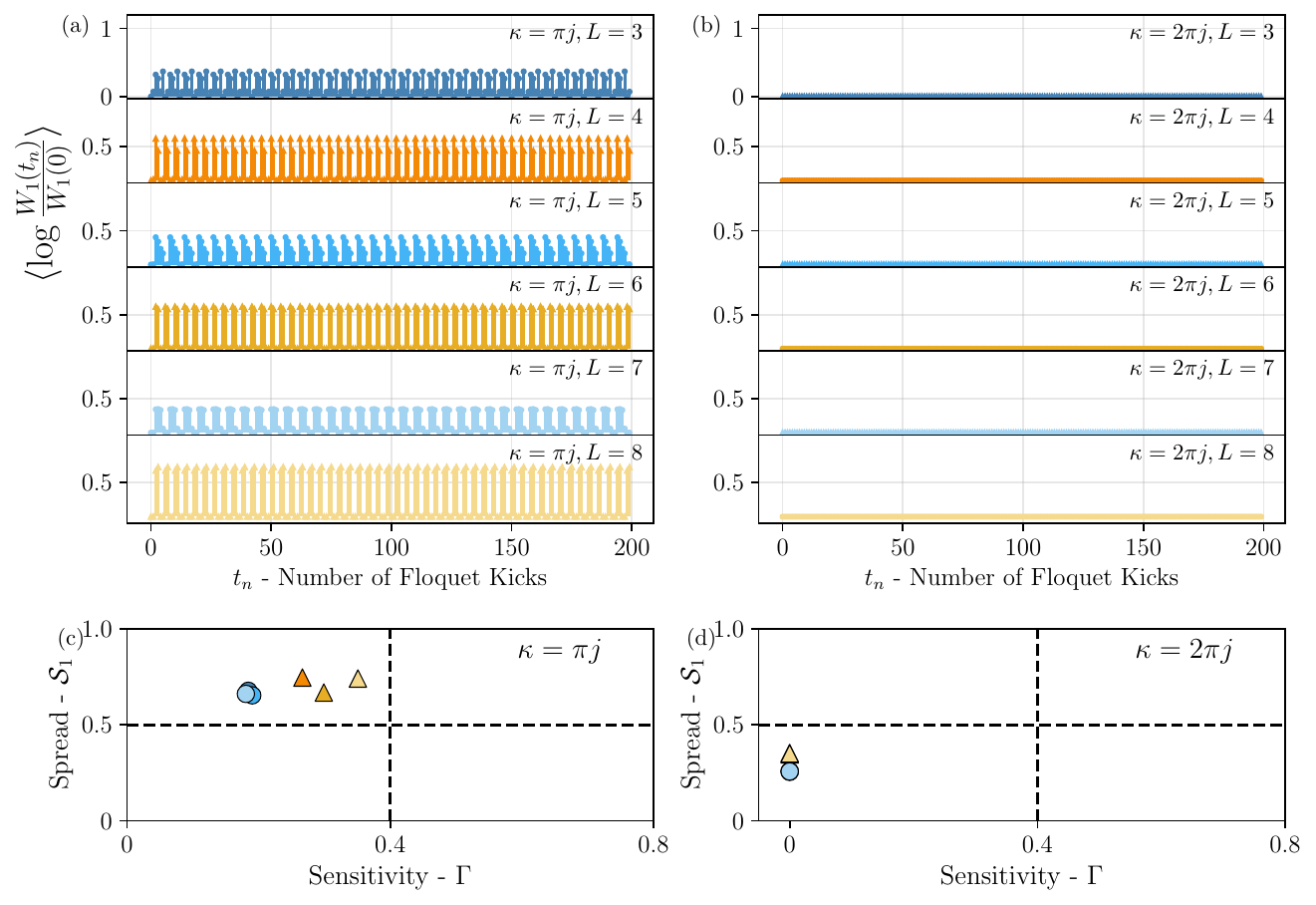}
\caption{
Separation dynamics and state-space exploration in recurrent kicked-top regimes for increasing system size $L$.
(a) For $\kappa=\pi j$, the average logarithmic relative separation
$\langle\ln[W_1(t_n)/W_1(0)]\rangle$ oscillates persistently, yielding
$\Gamma>0$ and indicating sensitivity to perturbations.
Integer-spin systems exhibit larger changes in distinguishability than
half-integer-spin systems.
(b) For $\kappa=2\pi j$, the distinguishability is preserved, giving $\Gamma=0$.
(c) The sensitivity--spread diagram for $\kappa=\pi j$ shows low sensitivity but relatively large coverage $\mathcal{S}_1$.
(d) For $\kappa=2\pi j$, both $\Gamma$ and $\mathcal{S}_1$ remain small, indicating localized recurrent dynamics with limited state-space exploration.
The dashed lines in panels (c) and (d) serve as visual guides separating the qualitative dynamical regimes and do not represent sharp thresholds.
}
    \label{fig:R1}
\end{figure*}

In contrast to the previous case, the separation between the two states is preserved throughout the evolution. The subsystem dynamics remain confined, similar to the noninteracting case $\kappa=0$, and the distinguishability measure vanishes, $\Gamma = 0$, indicating the absence of distinguishability growth.

Figure~\ref{fig:R1}(d) shows that both $\Gamma$ and $\mathcal{S}_1$ remain small for $\kappa=2\pi j$, placing these dynamics toward the localized, periodic part of the sensitivity--spread plane. 

Both behaviors follow from the structure of the kick. Since $L=2j$, the
nonlinear term contributes a factor
$e^{-i(\kappa/4j)\,\sigma_z^{(i)}\sigma_z^{(k)}}$ for each qubit pair. At
$\kappa=2\pi j$ this factor equals $-i\,\sigma_z^{(i)}\sigma_z^{(k)}$, a product
of single-qubit operators, so the Floquet operator acts locally and generates no
entanglement between $S$ and $E$. The GQS therefore remains a single point that
evolves under a single-qubit unitary, as in the noninteracting case, and
$\Gamma=0$. At $\kappa=\pi j$, by contrast, each pair factor
$e^{-i\pi\sigma_z^{(i)}\sigma_z^{(k)}/4}$ is maximally entangling. Entanglement
is generated and then undone within each recurrence period, so the projected
ensemble deforms in an initial-state-dependent way at intermediate times before
returning. Nearby ensembles separate during each cycle even though both
eventually recur, producing the oscillating distinguishability and positive
$\Gamma$ in Fig.~\ref{fig:R1}(a).

The comparison between $\kappa=\pi j$ and $\kappa=2\pi j$ shows that recurrence alone is not sufficient to characterize dynamical simplicity. What matters is whether recurrence preserves nearby-state distinguishability or allows sustained growth and broader state-space exploration.

\section{Phase-space structures for the \texorpdfstring{$L$}{L}-qubit kicked top}
\label{app:phase_space_structure}
To understand how the dynamical complexity of the projected ensemble depends on the initial condition, we compute the distinguishability measure $\Gamma$ and the state-space coverage index $\mathcal{S}_1$ over the spin-coherent state phase space of the $L$-qubit kicked top. For each initial condition $(\theta,\phi)$, the total system is initialized as a product spin-coherent state,
\begin{equation}
|\Psi_{SE}(0)\rangle
=
\bigotimes_{\ell=1}^{L}|\psi(\theta,\phi)\rangle .
\end{equation}
We then evolve the full $L$-qubit state unitarily, treat one qubit as the subsystem, and construct the corresponding geometric quantum state on the subsystem Bloch sphere. The quantities $\Gamma$ and $\mathcal{S}_1$ are evaluated on a $100\times100$ grid of initial conditions, producing phase-space maps that reveal how sensitivity and state-space exploration depend on the interaction strength $\kappa$ and the effective environment size $L-1$.

\begin{figure*}[htbp]
    \centering         
    \includegraphics[width=.95\linewidth]{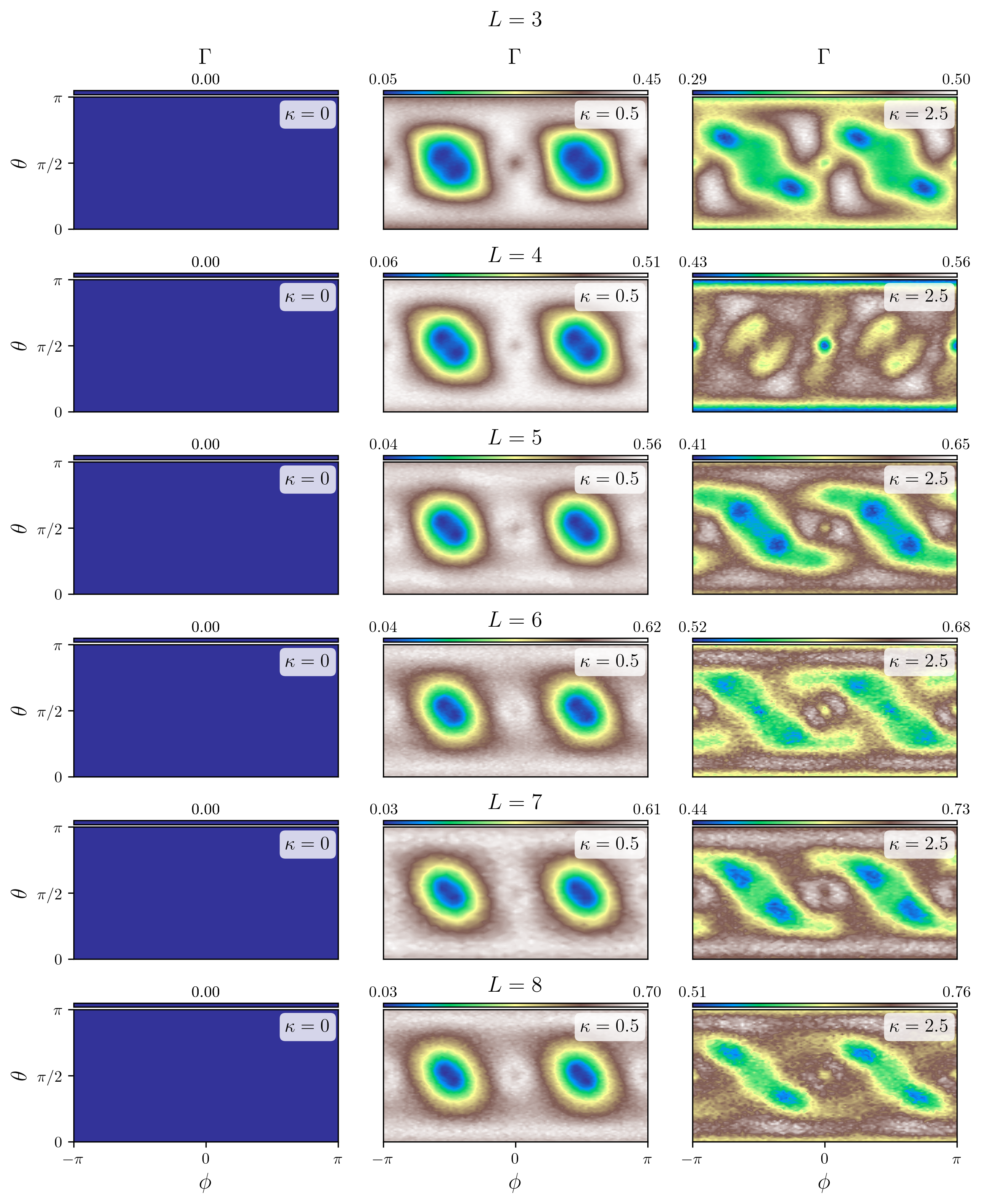}
    \caption{Phase-space structure of $\Gamma$ for the $L$-qubit kicked top. Columns show $\kappa=0,0.5,2.5$, and rows show $L=3,\ldots,8$. For $\kappa=0$, $\Gamma=0$ uniformly. Increasing $\kappa$ produces increasingly extended and intricate regions of distinguishability growth, with even-$L$ systems generally showing slightly stronger sensitivity than odd-$L$ systems.
    }
    \label{fig:Gamma_Lk}
\end{figure*}

Figure~\ref{fig:Gamma_Lk} shows the phase-space structure of the distinguishability measure $\Gamma$ for $L=3,\ldots,8$ and $\kappa=0,0.5,2.5$. In the noninteracting limit, $\kappa=0$, $\Gamma$ is uniformly zero for all $L$. Although the initial condition determines the subsystem trajectory, the
distance between two nearby states does not change with time, providing a baseline for periodic dynamics.

\begin{figure*}[htbp]
    \centering
     \includegraphics[width=0.95\linewidth]{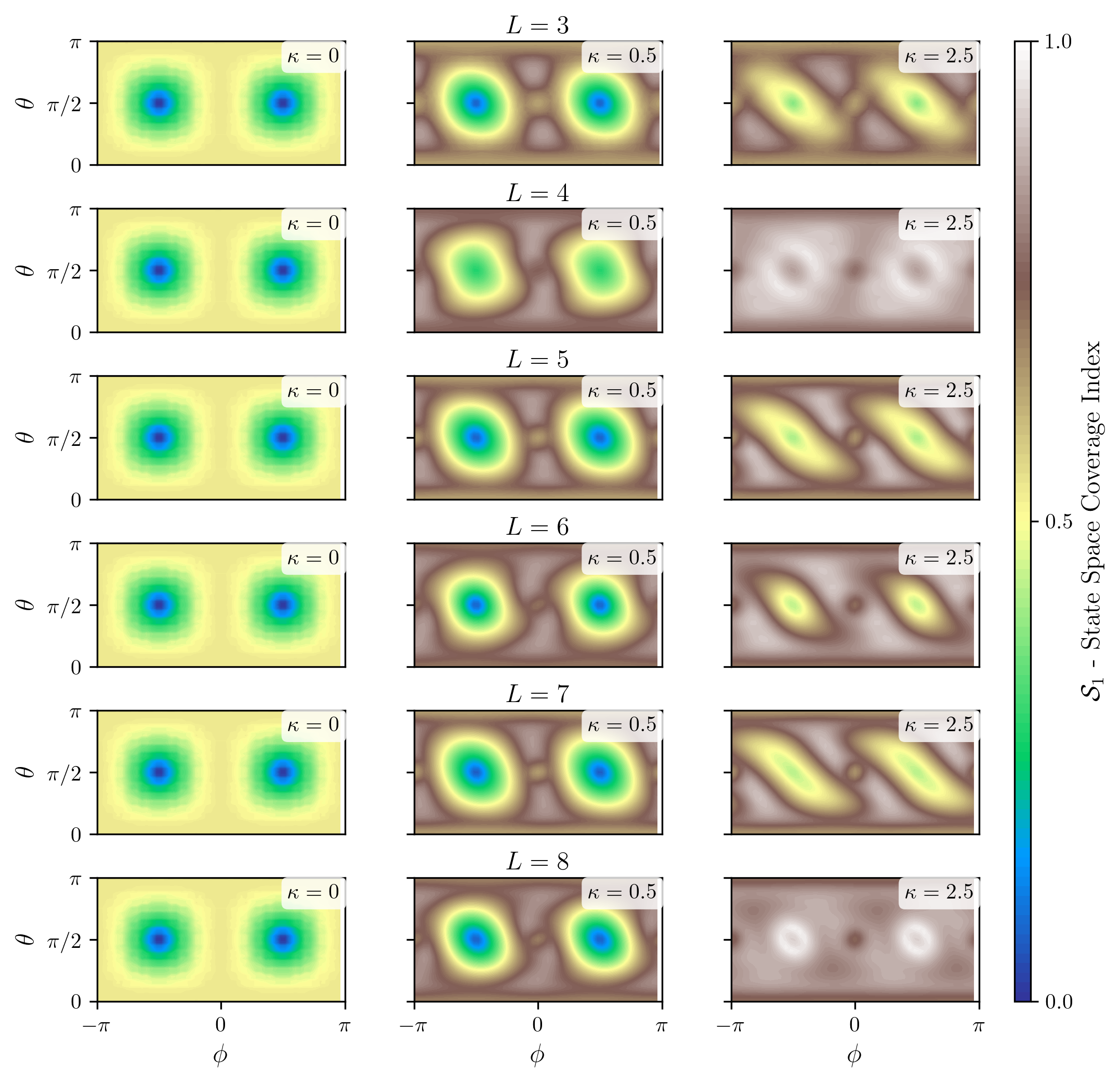}
    \caption{
    Phase-space structure of the coverage index $\mathcal{S}_1$ for the $L$-qubit kicked top.
    Columns show $\kappa=0,0.5,2.5$ and rows show $L=3,4,5,6,7,8$.
    Increasing $\kappa$ deforms the periodic coverage patterns and promotes broader exploration of the subsystem state space.
    Even-$L$ systems generally exhibit larger coverage than odd-$L$ systems, indicating parity-dependent differences in interaction-induced spreading.
    }
    \label{fig:coverage_Lk}
\end{figure*}

At $\kappa=0.5$, regions of enhanced $\Gamma$ emerge, showing that sensitivity depends strongly on the initial spin-coherent state. Some initial conditions remain weakly sensitive, while others generate larger distinguishability growth through interaction-induced entanglement with the remaining qubits.

For stronger interactions, $\kappa=2.5$, the high-$\Gamma$ regions broaden and become more intricate. Sensitivity increases over a larger portion of phase space, indicating that stronger subsystem-environment interactions promote more widespread distinguishability growth. The maps also reveal finite-size parity-symmetry effects, with even-$L$ systems (integer total spin) generally exhibiting larger $\Gamma$ values and stronger phase-space deformation than odd-$L$ systems (half-integer total spin).

Figure~\ref{fig:coverage_Lk} shows the corresponding phase-space structure of the state-space coverage index $\mathcal{S}_1$. While $\Gamma$ measures distinguishability growth between nearby GQSs, $\mathcal{S}_1$ quantifies the long-time extent of subsystem exploration over the Bloch sphere. The two diagnostics therefore capture complementary aspects of the dynamics, namely finite-time sensitivity and global state-space spreading.

For $\kappa=0$, the coverage maps retain patterns reflecting periodic dynamics and are strongly organized by the initial spin-coherent state. Unlike $\Gamma$, which is uniformly zero in this limit, $\mathcal{S}_1$ varies across phase space because it is set by the geometry of the trajectory. States near $(\theta,\phi)=(\pi/2,\pi/2)$, an eigenstate of the non-interacting portion of the Hamiltonian $H_0 = (\pi/2\tau)J_y$, remain confined and yield small $\mathcal{S}_1$, whereas states farther away trace larger orbits and produce greater long-time coverage.

As $\kappa$ increases to $0.5$, the coverage patterns broaden and deform.
At \(\kappa=2.5\), the coverage becomes larger and more extended, with many regions reaching $\mathcal{S}_1>0.5$. Stronger interactions therefore promote global exploration while preserving the underlying phase-space organization.

Parity-symmetry effects are more visible in the coverage maps than in the sensitivity maps. Even-$L$ systems generally show broader and more uniform coverage than odd-$L$ systems, indicating differences between integer-spin and half-integer-spin dynamics that are sometimes more evident in $\mathcal{S}_1$ than in $\Gamma$.

Figures~\ref{fig:Gamma_Lk} and~\ref{fig:coverage_Lk} show that interaction strength and environment size reshape the phase-space structure of the projected ensemble in complementary ways. Increasing $\kappa$ enhances both distinguishability growth and long-time coverage, while increasing $L$ reveals systematic parity effects. The resulting patterns also resemble the classical kicked-top structures in the maximal Lyapunov exponent maps of Fig.~\ref{fig:F4}(c,d), where regions of enhanced quantum distinguishability and coverage appear in organized bands and islands rather than uniformly. This suggests that the geometry of the underlying kicked-top phase space continues to organize the dynamical complexity of the projected ensemble, even though the subsystem dynamics are described by probability measures on quantum state space rather than classical trajectories.

\end{document}